\documentclass[preprint,showpacs,amsmath,amssymb,aps,prd,nofootinbib]{revtex4}
\usepackage{epsfig,graphicx,color,appendix,fge}
\begin{document}
\begin{flushright}
CAU-THEP-19-07
\end{flushright}
\def\CP{{\it CP}~}
\def\cp{{\it CP}}
\title{\mbox{}\\[10pt] 
A Model for Neutrino Anomalies and IceCube data}

\author{Y. H. Ahn}
\affiliation{Department of Physics, Chung-Ang University, Seoul 06974, Korea.}
\email{axionahn@naver.com}
\author{Sin Kyu Kang}
\affiliation{School of Liberal Arts, Seoul Tech, Seoul, 01811 Korea}
\email{skkang@seoultech.ac.kr}




\begin{abstract}
We interpret the neutrino anomalies in neutrino oscillation experiments and the high energy neutrino events at IceCube in terms of neutrino oscillations in an extension of the standard model where three sterile neutrinos are introduced so as to make
two light neutrinos to be Pseudo-Dirac particles and a light neutrino to be a Majorana particle. 
Our model is different from the so-called $3+n$ model with $n$ sterile neutrinos suggested to interpret short baseline anomalies in terms of neutrino oscillations.
While the Pontecorvo-Maki-Nakagawa-Sakata (PMNS) matrix in $3+n$ model is simply extended to $n\times n$ unitary matrix,  the neutrino mixing matrix in our model is parameterized so as to keep the $3\times3$ PMNS mixing matrix for three active neutrinos unitary.
There are also no flavor changing neutral current interactions leading to the conversion of active neutrinos to sterile ones or vice versa.
We derive new forms of neutrino oscillation probabilities containing the new interference between the active and sterile neutrinos which are characterized by additional new parameters $\Delta m^2$ and $\theta$.
Based on the new formulae derived, we show how the short baseline neutrino anomalies can be explained in terms of oscillations, and  study the implication of the high energy neutrino events detected at IceCube on the probe of pseudo-Dirac neutrinos.
New phenomenological effects attributed to the existence of the sterile neutrinos are discussed. 
%

\end{abstract}

\maketitle %
\section{Introduction}
The observation of neutrino oscillations in the atmospheric neutrino\,\footnote{It has been confirmed by the K2K\,\cite{Ahn:2004te} and MINOS\,\cite{Adamson:2005qc} accelerator based experiments.} Super-Kamiokande\,\cite{Fukuda:1998mi},  the solar neutrino\,\footnote{It has been confirmed by the reactor neutrino experiment KamLAND\,\cite{Araki:2004mb}} SNO\,\cite{Ahmad:2002jz}, and  the reactor neutrino  Daya Bay\cite{An:2012eh} and RENO\,\cite{Ahn:2012nd}\,\footnote{See also Double Chooz\,\cite{Abe:2011fz}.} experiments is one of big discoveries in particle physics since 90's. 
It implies that neutrinos are massive particles and that
the three flavor neutrinos $\nu_e, \nu_\mu, \nu_\tau$ are mixtures of neutrinos with definite masses $\nu_i$ (with $i=1,2,...$).
Although neutrino oscillations among three active neutrino flavors have been confirmed by the analysis based on the experiments mentioned above, there exist several anomalies which are unexpected results coming from short baseline (SBL) experiments such as 
the reactor antineutrino anomaly\,\cite{Mention:2011rk}, the Gallium solar anomaly\,\cite{gallium1, gallium2}, and the Liquid Scintillator Neutrino Detector (LSND) anomaly\,\cite{Athanassopoulos:1995iw} (including MiniBooNE anomaly\,\cite{Aguilar-Arevalo:2018gpe}). 
To resolve those neutrino anomalies in terms of neutrino oscillation, it is required to introduce
at least one additional squared-mass difference, $\Delta m^2_{\rm SBL}$, which is much larger than $\Delta m^2_{\rm Sol}$ and $\Delta m^2_{\rm Atm}$ \cite{Mention:2011rk}. 
This result suggests indication in favor of the possible existence of eV-mass sterile neutrino.


Apart from the anomalous results in accelerator and reactor based neutrino experiments favoring the existence of light sterile neutrino, IceCube experiments\,\cite{icecube} announced the observation of vey high energy neutrino events.
The study for the track-to shower ratio of the subset with energy above 60 TeV coming from IceCube has shown that the events are consistent with the hypothesis that cosmic neutrinos have been seen even though their origin and propagation are still elusive \cite{Palladino:2015zua}.
In order to examine them, we need to discriminate the flavor composition of cosmic neutrinos, which is possible by looking at the topology of the events. But, the current limited statistics does not allow yet to discriminate the initial flavor.
It is also widely perceived that the IceCube may serve as 
astronomical-scale baseline experiment to uncover the oscillation effects due to very tiny mass splitting of the pseudo-Dirac neutrinos.
If the oscillation effects induced by pseudo-Dirac neutrinos with very high energy and long trajectory are prominent, then they may affect the
observables such as the neutrino flavor composition detected from the ultra-high energy neutrino experiments.
Since the current precision on the observation of very high energy neutrino events does not exclude such an oscillation effect, it would be meaningful to confront  any model for pseudo-Dirac neutrinos realized by introducing sterile neutrinos with the high energy astrophysical neutrino events.

In this work, we construct a model having three sterile neutrinos which make two light neutrinos to be pseudo-Dirac particles and a light neutrino to be Majorana particle.
Then, we investigate if the sterile neutrino that makes a light neutrino to be Majorana particle can play a crucial role in resolving the neutrino anomalies from SBL experiments, and study the implication of the very high energy astrophysical neutrino data  from IceCube on the probe of the  pseudo-Dirac neutrinos.
%
%
Thus, the goal of this paper is to interpret  both SBL neutrino anomalies and astronomical neutrino data observed at IceCube
in terms of neutrino oscillations\,\cite{Ahn:2016hbn, Ahn:2016hhq, ice_ref} in the context of the model we construct.

Our model is different from the so-called $3+n$ model with $n$ sterile neutrinos suggested to interpret SBL anomalies in terms of neutrino oscillations. 
While the Pontecorvo-Maki-Nakagawa-Sakata (PMNS) matrix in $3+n$ model is simply extended to $n\times n$ unitary matrix as in Refs.\,\cite{Mohapatra:2005wk, white, Gariazzo:2015rra},  the neutrino mixing matrix in our model is 
parameterized so as to keep the $3\times3$ PMNS mixing matrix for three active neutrinos
unitary.
In this model, there are no flavor changing neutral current interactions leading to the conversion of active neutrinos to sterile ones or vice versa.
We will present new forms of neutrino oscillation probabilities modified by introducing new sterile neutrinos.
The interference between active flavor and sterile neutrinos due to new additional oscillation parameters $\Delta m^2$ and mixing angle $\theta$ triggers new oscillation effects which can be responsible for the explanation of SBL neutrino anomalies and very high energy neutrino events at IceCube.
Based on the new formulae for neutrino oscillations, we will discuss how SBL neutrino anomalies can be
explained or alleviated and consistently accommodate the recent  IceCube high energy neutrino events.
Constraints on the oscillation parameters  coming from solar and atmospheric neutrino data, cosmological
observation for the sum of active neutrino masses, effective neutrino mass in $\beta$-decay and neutrinoless-double-beta ($0\nu\beta\beta$)-decay experiments will be discussed.
Our study based on the terrestrial, atmospheric and solar experiments is similar to that in the so-called $3+1$ model in the light that the two
very tiny pseudo-Dirac mass splittings are not relevant for those experiments. However, this work includes the study for the  implication of IceCube data
on the probe of the oscillarion effects induced by those two tiny mass splittings while keeping the results for the explanations of SBL neutrino anomalies,
atmopsheric and solar neutrino experiments in terms of neutrino oscillations including an eV scale sterile neutrino.

This work is organized as follows. In section II, we discuss our model and study neutrino masses and mixings in the new framework. In section III, we study how the new parameters could be constrained through the cosmological data (the sum of active neutrino masses), and the effective neutrino masses in $\beta$-decay and $0\nu\beta\beta$-decay experiments. In section I,V we develop the new active neutrino oscillation probabilities modified by incorporating new sterile neutrinos.
And we study how SBL neutrino anomalies can be explained. In section V, we investigate how new effects
due to sterile neutrino in the solar and atmospheric oscillations
can be  constrained by the SBL $\nu_e(\bar{\nu}_e)$ disappearance and $\nu_\mu(\bar{\nu}_\mu)$ disappearance channels, and  examine whether the high energy neutrino events from IceCube data can
be interpreted in terms of neutrino oscillation. In section VI we examine astronomical neutrino data observed at IceCube to uncover the oscillation effects of tiny mass splittings.
In section VII, we state conclusions by summarizing this work.
\section{Masses and Mixings}
Introducing  right-handed singlet  neutrinos $N_R$,  extra neutrino singlet fermions $S$, and 
an SU(2)$_L$ singlet scalar field $\Psi$.
we construct the renormalizable Lagrangian given in the charged lepton basis as \cite{Ahn:2016hhq}
\begin{eqnarray}
-{\cal L}=\frac{1}{2}\overline{N_R^c}\,M_{R}\,N_R+ \overline{L}\,\tilde{\Phi}Y_{D}\,N_R+ \overline{L}\,\tilde{\Phi}\,Y_{DS}\,S  +
 \overline{S^c}\,\Psi\,Y_{S}\,N_R +\frac{1}{2} \overline{S^c}\,\mu\,S +h.c.~,
\label{Lag}
\end{eqnarray}
where $N_R, S$ are three generations, $L$ stand for SU(2)$_L$ left-handed lepton doublet, $\Phi=(\phi^{+}, \phi^{0})^T$ is the SM Higgs doublet and $\tilde{\Phi}\equiv i\tau_{2}\Phi^{\ast}$.
$M_{R}$ and $\mu$ are Majorana masses for the $N_R$ and $S$ fields, respectively.
The above  Lagrangian is invariant under $U(1)_{B-L}$ when $\mu,M_R=0$  by assigning
 quantum numbers $L:1$, $N_{R}, S:1$, $\Psi:-2$, and $\Phi:0$ under the $U(1)_L$ (or $U(1)_{B-L}$) symmetry. 
Then, the parameters $\mu$ and $M_R$ reflect soft symmetry breaking of $U(1)_L$.
Nontrivial vacuum expectation value (VEV) of the scalar field $\Psi$  does not break the electroweak symmetry, but spontaneously breaks the $U(1)_L$ (or $U(1)_{B-L}$) symmetry.
Thus the symmetry breaking scale for $\Psi$  can be different from the electroweak scale.
Integrating out the heavy Majorana neutrinos in the Lagrangian Eq. (\ref{Lag}),  we obtain effective Lagrangian for neutrino sectors given by,
\begin{eqnarray}
-{\cal L}_{\rm eff}&=& \overline{\nu_L}\,\phi^{0}\,Y_{DS}\,S -\frac{1}{2}\overline{\nu_L}\,\phi^{0}\,Y_{D}M^{-1}_{R}Y^{T}_{D}\,\phi^{0}\,\nu^c_L-
 \overline{\nu_L}\,\phi^{0}\,Y_{D}M^{-1}_{R}Y^{T}_{S}\,\Psi\,S\nonumber \\
& & -\frac{1}{2}\overline{S^c}\,\Psi\,Y_SM^{-1}_{R}Y^{T}_{S}\,\Psi\,S
 + \frac{1}{2} \overline{S^c}\,\mu\,S+h.c.~,
\end{eqnarray}
where $Y_{D}, Y_{S}, Y_{DS}, M_{R}$ and $\mu$ are all $3\times3$ matrices.

After the scalar fields $\Phi$ and $\Psi$ get VEVs and taking $S$ to be right-handed,
the Lagrangian for neutrinos in the charged lepton basis reads
 \begin{eqnarray}
-{\cal L}_{\nu} &=& 
 \frac{1}{2} \begin{pmatrix} \overline{\nu^c_L} & \overline{S_R} \end{pmatrix} {\cal M}_{\nu} \begin{pmatrix} \nu_L \\ S^c_R  \end{pmatrix} +\frac{g}{\sqrt{2}}W^-_\mu\overline{\ell_L}\gamma^\mu\,\nu_L+\text{h.c.}+\frac{g}{2\cos\theta_W}Z_\mu\overline{\nu_L}\gamma^\mu\,\nu_L\,,
  \label{lag}
 \end{eqnarray}
 where $g$ is the $SU(2)$ coupling constant, $\theta_W$ is the Weinberg angle, $\ell=(e, \mu, \tau)$, $\nu_L=(\nu_e, \nu_\mu, \nu_\tau)$, and $S_R=(S_1, S_2,...S_n)$. The light neutral fermions $S_\alpha$ do not take part in the standard weak interaction and thus are not excluded by LEP results, while the number of active neutrinos coupled with the $W^{\pm}$ and $Z$ bosons is $N_{\nu}=2.984\pm0.008$\,\cite{PDG}.
After electroweak symmetry breaking, Eq.\,(\ref{lag}) describes $3+ n$ Majorana neutrinos. In the case of $n=3$ sterile neutrinos, the $6\times6$ Majorana neutrino mass matrix is
 \begin{eqnarray}
{\cal M}_{\nu} = \begin{pmatrix} M_L & M^T_{D}  \\ M_{D} & M_{S}  \end{pmatrix} \,,
  \label{nu_matr}
 \end{eqnarray}
where the $3\times3$ mass matrices $M_{D}$, $M_L$, and $M_S$ are those for Dirac masses, left- and right-handed Majorana masses, respectively, given by
\begin{eqnarray}
M_L&=&-m_{D}M^{-1}_{R}m^{T}_{D}, \nonumber \\
M_D&=&m_{DS}-m_{D}M^{-1}_{R}m^{T}_{S}, \nonumber \\
M_S &=& \mu-m_{S}M^{-1}_{R}m^{T}_{S},
\label{massEle}
\end{eqnarray}
where $m_{D}=Y_{D}\langle\phi^{0}\rangle, m_{S}=Y_{S}\langle\Psi\rangle$ and $m_{DS}=Y_{DS}\langle\phi^{0}\rangle$.
Here we take $M_R \gg m_S \simeq m_{D}\gg \mu$, and  neutrinos become pseudo-Dirac particles when  $M_D$ is dominant over $M_L$ and
$M_S$ in Eq. (\ref{nu_matr}), which reflects $m_{DS} \gg (m_D m_S)/M_R$.
In order to get physical parameters, we perform basis rotations from interaction eigenstates to mass eigenstates\,\cite{Kobayashi:2000md, Ahn:2016hbn},
 \begin{eqnarray}
 \begin{pmatrix} \nu_L \\ S^c_R  \end{pmatrix} \rightarrow W^\dag_\nu\begin{pmatrix} \nu_L \\ S^c_R  \end{pmatrix}\equiv\xi_L
  \label{Wnu0}
 \end{eqnarray}
where $\xi_{L}$ is the mass eigenstate of neutrino, Here, the $6\times6$ unitary neutrino transformation matrix $W_\nu$ given by, 
 \begin{eqnarray}
 W_\nu&=& {\left(\begin{array}{cc}
 U_L & 0_3  \\
 0_3 &  U_R 
 \end{array}\right)}{\left(\begin{array}{cc}
 V_1 & iV_1  \\
 V_2 &  -iV_2 
 \end{array}\right)}V_\nu \nonumber \\
  \label{Wnu}
\end{eqnarray}
 where
  \begin{eqnarray}
V_\nu={\left(\begin{array}{cccccc}
 e^{i\phi_1}\cos\theta_1 & 0 & 0 &  -e^{i\phi_1}\sin\theta_1 & 0 & 0  \\
  0 & e^{i\phi_2}\cos\theta_2 & 0 & 0 &  -e^{i\phi_2}\sin\theta_2 & 0   \\
   0 & 0 & e^{i\phi_3}\cos\theta_3 & 0 & 0  &  -e^{i\phi_3}\sin\theta_3   \\
e^{-i\phi_1} \sin\theta_1 & 0 & 0&  e^{-i\phi_1}\cos\theta_1 & 0 & 0 \\
0 & e^{-i\phi_2} \sin\theta_2 & 0 & 0&  e^{-i\phi_2}\cos\theta_2 & 0  \\
0 & 0 & e^{-i\phi_3} \sin\theta_3 & 0 & 0&  e^{-i\phi_2}\cos\theta_3   
 \end{array}\right)}\,.
 \end{eqnarray}
In the above expression, $0_3$ is the $3\times3$ null matrix, $U_R$ is an unknown $3\times3$ unitary matrix, $V_1={\rm diag}(1,1,1)/\sqrt{2}$, $V_2={\rm diag}(e^{i\varphi_1}, e^{i\varphi_2}, e^{i\varphi_3})/\sqrt{2}$ with $\varphi_i$ being arbitrary phases.
The matrices $V_1$ and $V_2$ are presonsible for the maximal mixing between active neutrinos and sterile neutrinos.
The angle  $\theta_k$ is introduced thanks to nondegeneracy between $M_L$ and $M_S$ in eq.(\ref{nu_matr}), and thus it is responsible for the deviation of maximal mixing between active neutrino and sterile neutrino, which reflects the breaking of degeneracy of a pair of neutrinos in each generations.
It is easily see that maximal mixing between $\nu_k$ and $S^c_k$ is recovered in the limit of $\theta_k=0$.
 We also note that the $3\times 3$ unitary matrix $U_L$ should be  the PMNS neutrino mixing matrix
responsible for the mixing among three active neutrinos.
Then the neutrino mass matrix ${\cal M}_{\nu}$ is  diagonalized in the mass eigenstates $(\nu_1, \nu_2, \nu_3, S^c_1, S^c_2, S^c_3)$ basis as 
 \begin{eqnarray}
  W^T_\nu{\cal M}_\nu W_\nu=V^T_\nu{\left(\begin{array}{cc}
 \hat{M}_{L} &  \hat{M}  \\
 \hat{M} &  \hat{M}_S 
 \end{array}\right)}V_\nu
 \equiv{\rm diag}(m_{\nu_1},m_{\nu_2},m_{\nu_3},m_{s_1},m_{s_2},m_{s_3})\,.
  \label{diag1}
 \end{eqnarray}
The expression Eq.\,(\ref{diag1}) represents that the Majorana mass matrices $M_{L}$ and $M_S$, and Dirac neutrino mass matrix, $M_{D}$ in Eq.(\ref{nu_matr}) are diagonalized by the mixing matrices $U_L$ and $U_R$ as $\hat{M}_{L}=U^T_LM_{L}U_L$, $\hat{M}_{S}=U^T_RM_{S}U_R$, and $\hat{M}=U^T_R\,M_{D}\,U_L={\rm diag}(m_{1},m_{2},m_{3})$.
To get the real and positive mass squared for the 6 neutrino mass eigenstates, we diagonalize
 the Hermitian matrix ${\cal M}_\nu{\cal M}^\dag_\nu$ with the help of Eq.\,(\ref{diag1}) as follows,
 \begin{eqnarray}
   &W^T_\nu\,{\cal M}_\nu{\cal M}^\dag_\nu\,W^\ast_\nu=\nonumber\\
   & V^T_\nu{\left(\begin{array}{cc}
 |\hat{M}|^2+|\hat{M}||\delta|+\frac{1}{2}(|\hat{M}_L|^2+|\hat{M}_S|^2) &  -\frac{i}{2}(|\hat{M}_L|^2-|\hat{M}_S|^2)  \\
 \frac{i}{2}(|\hat{M}_L|^2-|\hat{M}_S|^2) &  |\hat{M}|^2-|\hat{M}||\delta|+\frac{1}{2}(|\hat{M}_L|^2+|\hat{M}_S|^2) 
 \end{array}\right)}V^\ast_\nu\,,
  \label{eff_nu_mass}
 \end{eqnarray}
 where $\delta$ stands for $\delta_k\equiv(\hat{M}_{L})_k+(\hat{M}^\dag_S)_k$ originated from the left- and right-Majorana masses ($k=1,2,3$).
 From the above equation the mixing parameters $\theta_k$ and $\phi_k$ in Eq.\,(\ref{Wnu}) can be obtained
 \begin{eqnarray}
  \tan2\theta_k=\frac{|(\hat{M}_L)_k|^2-|(\hat{M}_{S})_k|^2}{2\hat{M}_k|\delta_k|}\quad\text{and}~~ \phi_k=\frac{\pi}{4}\,.
 \end{eqnarray}

Now, to accommodate both an eV sterile neutrino for a possible solution to the neutrino anomalies\,\cite{Mention:2011rk, Athanassopoulos:1995iw, Aguilar-Arevalo:2018gpe} and the high energy neutrino events in the IceCube detector\,\cite{icecube} to be interpreted as new neutrino oscillations, simultaneously, we assume that $m_{\nu_3}\ll m_{s_3}$ and $m_{\nu_1}\approx m_{s_1}$, $m_{\nu_2}\approx m_{s_2}$,\,\footnote{This possibility could theoretically be realized in a non-renomalizable flavor model considering non-Abelian discrete symmetry  plus Abelian symmetry, e.g. Refs.\,\cite{Ahn:2018cau, Ahn:2016hbn} where light active neutrino masses are mainly generated by QCD anomalous $U(1)$ symmetry (via Froggatt-Nielson mechanism\,\cite{Froggatt:1978nt}) while leptonic mixing matrix is produced (through seesaw formula\,\cite{Minkowski:1977sc}) by non-Abelian discrete symmetry.}.
It is equivalent to take the limit of both $\hat{M}_{j}\gg|(\hat{M}_{L})_j|\gg|(\hat{M}_{S})_j|$ (with $j=1,2$) 
and $|(\hat{M}_{S})_3|\gg|(\hat{M}_{L})_3|$, leading to
\footnote{The Dirac masses of first and second generations are much larger than the left(right)-handed Majorana masses of those, while the Dirac mass of third generation is larger than the left(right)-handed Majorana masses of that for $3\pi/8<\theta_3<\pi/2$ and smaller for $\pi/4<\theta_3\leq3\pi/8$. For $|(\hat{M}_{L})_3|\gg|(\hat{M}_{S})_3|$ is not realized due to a requirement $m^2_{s_3}\gg m^2_{\nu_3}$. See  Eq.\,(\ref{msd}).}
 \begin{eqnarray}
   \delta_{1(2)}\simeq(\hat{M}_{L})_{1(2)}\qquad\text{and}\qquad|\delta_{3}|\simeq-2m_3\tan2\theta_3
  \label{angle00}
 \end{eqnarray}
which in turn gives $\theta_{1}\approx\theta_2\approx0$ and $\pi/4<\theta_3<\pi/2$. Since the active neutrinos are massive and mixed, the weak eigenstates $\nu_{\alpha}$ (with flavor $\alpha=e,\mu,\tau$) produced in a weak gauge interaction are linear combinations of the mass eigenstates with definite masses. The three neutrino active states emitted by weak interactions are described in terms of the mass eigenstates $\xi_k=(\nu_k$\, $S^c_k)$ ($k=1,2,3$) as
 \begin{eqnarray}
   \nu_\alpha=\sum_{k=1}^3U_{\alpha k}\,\xi_k
  \label{nu_mass_eigen}
 \end{eqnarray} 
with
 \begin{eqnarray}
  \xi_{k=1,2}=\frac{1}{\sqrt{2}}\left(\begin{array}{cc}
 1 & i \end{array}\right)\left(\begin{array}{c}
 \nu_{k} \\
 S^c_k \end{array}\right)~\text{and}~~\xi_{3}=\frac{1}{\sqrt{2}}\left(\begin{array}{cc}
 \cos\theta_3+\sin\theta_3 & \cos\theta_3-\sin\theta_3  \end{array}\right)\left(\begin{array}{c}
 \nu_{3} \\
 S^c_3 \end{array}\right)\,,
  \label{nu_mass_eigen2}
 \end{eqnarray}
in which the field redefinitions $\nu_j\rightarrow e^{i\frac{\pi}{4}}\nu_j$, $S^c_j\rightarrow e^{-i\frac{\pi}{4}}S^c_j$ (with $j=1,2$) and $\nu_3\rightarrow e^{i\frac{\pi}{4}}\nu_3$, $S^c_3\rightarrow e^{i\frac{\pi}{4}}S^c_3$ are used. In Eq.\,(\ref{nu_mass_eigen}), $U$ is the $3\times3$ PMNS mixing matrix $U_{\rm PMNS}$ which is expressed in terms of three mixing angles, $\theta_{12}, \theta_{13}, \theta_{23}$, and three \cp-odd phases (one $\delta_{CP}$ for the Dirac neutrino and two $\varphi_{1,2}$ for the Majorana neutrino) as\,\cite{PDG}
 \begin{eqnarray}
  U_{\rm PMNS}=
  {\left(\begin{array}{ccc}
   c_{13}c_{12} & c_{13}s_{12} & s_{13}e^{-i\delta_{CP}} \\
   -c_{23}s_{12}-s_{23}c_{12}s_{13}e^{i\delta_{CP}} & c_{23}c_{12}-s_{23}s_{12}s_{13}e^{i\delta_{CP}} & s_{23}c_{13}  \\
   s_{23}s_{12}-c_{23}c_{12}s_{13}e^{i\delta_{CP}} & -s_{23}c_{12}-c_{23}s_{12}s_{13}e^{i\delta_{CP}} & c_{23}c_{13}
   \end{array}\right)}P_{\nu}~,
 \label{PMNS}
 \end{eqnarray}
where $s_{ij}\equiv \sin\theta_{ij}$, $c_{ij}\equiv \cos\theta_{ij}$ and $P_{\nu}$ is a diagonal phase matrix what is that particles are Majorana ones.

And their mass eigenvalues (real and positive) are given as
\begin{eqnarray}
 m^2_{\nu_j}&=&m_j^2+m_j|\delta_j|+\frac{1}{2}(|(\hat{M}_L)_j|^2+|(\hat{M}_S)_j|^2)\,,\nonumber\\
 m^2_{\nu_3}&=& m_3^2+\frac{1}{2}(|(\hat{M}_L)_3|^2+|(\hat{M}_S)_3|^2)+\frac{m_3|\delta_3|}{\cos2\theta_3}\,,\nonumber\\
  m^2_{s_j}&=&m_j^2-m_j|\delta_j|+\frac{1}{2}(|(\hat{M}_L)_j|^2+|(\hat{M}_S)_j|^2)\,,\nonumber\\
   m^2_{s_3}&=& m_3^2+\frac{1}{2}(|(\hat{M}_L)_3|^2+|(\hat{M}_S)_3|^2)-\frac{m_3|\delta_3|}{\cos2\theta_3}\,.
  \label{nu_mass}
\end{eqnarray}
The neutrino masses for the first and second generations lift slightly the degeneracy of mass-eigenvalues, and we get almost degenerate pairs of eigenstates with tiny mass differences: the mass-squared differences in each pair $\Delta m^2_{k}\equiv m^2_{\nu_k}-m^2_{s_k}$ (with $k=1,2$) are so small that the same mass ordering should apply to both eigenmasses, that is, 
 \begin{eqnarray}
   \Delta m^2_{k}=2m_k|\delta_k|\ll m^2_{\nu_k}\quad\text{for}~k=1,2\,.
  \label{msd}
 \end{eqnarray}
On the other hand, the mass splitting for third generation is given by
 \begin{eqnarray}
   \Delta m^2_{3}\equiv m^2_{\nu_3}-m^2_{s_3}=2\frac{m_3|\delta_3|}{\cos2\theta_3}\,,
  \label{msd}
 \end{eqnarray}
leading to $-1<\cos2\theta_3<0$, that is, $\pi/4<\theta_3<3\pi/4$ due to the requirement of $m^2_{s_3}\gg m^2_{\nu_3}$. 
From Eqs.\,(\ref{angle00}) and (\ref{msd}) a possible range of $\theta_3$ can be derived as
 \begin{eqnarray}
   \frac{\pi}{4}<\theta_3<\frac{\pi}{2}\,,
  \label{angle}
 \end{eqnarray}
in which, especially, for $3\pi/8<\theta_3<\pi/2$ the third neutrino pair could be Majorana.
As is well-known, because of the observed hierarchy $|\Delta m^{2}_{\rm Atm}|= |m^{2}_{\nu_3}-(m^{2}_{\nu_1}+m^{2}_{\nu_2})/2|\gg\Delta m^{2}_{\rm Sol}\equiv m^{2}_{\nu_2}-m^{2}_{\nu_1}>0$, and the requirement of a Mikheyev-Smirnov-Wolfenstein (MSW) resonance for solar neutrinos\,\cite{Wolfenstein:1977ue}, there are two possible neutrino mass spectra: (i) the normal mass ordering (NO) $m^2_{\nu_1}\approx m^2_{s_1}<m^2_{\nu_2}\approx m^2_{s_2}<m^2_{\nu_3}\ll m^2_{s_3}$, and (ii) the inverted mass ordering (IO) $m^2_{\nu_3}<m^2_{\nu_1}\approx m^2_{s_1}<m^2_{\nu_2}\approx m^2_{s_2}\ll m^2_{s_3}$. 
We use the following global fit values and $3\sigma$ intervals for physics parameters
 \begin{eqnarray}
  &\Delta m^2_{21}=7.55^{+0.59}_{-0.50}\times10^{-5}\,{\rm eV}^2\,,\qquad\qquad\theta_{12}[^\circ]=34.5^{+3.5}_{-3.0}\,,\nonumber\\
  &|\Delta m^2_{31}|=2.50^{+0.10}_{-0.09}\times10^{-3}\,(2.42^{+0.09}_{-0.11}\times10^{-3})\,{\rm eV}^2\,,\qquad\quad\theta_{23}[^\circ]=47.7^{+3.0}_{-5.9}\,(47.9^{+2.8}_{-5.6})\,,\nonumber\\
  &\theta_{13}[^\circ]=8.45^{+0.45}_{-0.45}\,(8.53^{+0.47}_{-0.43})\,,\qquad\qquad\qquad\delta_{CP}[^\circ]=238^{+111}_{-81}\,(281^{+68}_{-79})\,,
 \label{mixing_angle}
 \end{eqnarray}
where $\Delta m^2_{kj}\equiv m^2_{\nu_k}-m^2_{\nu_j}$, for normal mass ordering (inverted mass ordering) respectively\,\cite{deSalas:2017kay}.
Using above Eq.\,(\ref{angle00}), we can obtain  good approximated forms of $m^2_{\nu_3}$ and $ m^2_{s_3}$ in terms of $m_3$ and $\theta_3$ given as
\begin{eqnarray}
 m^2_{\nu_3}\approx m_3^2\,\frac{(1-\sin2\theta_3)^2}{\cos^22\theta_3}\,,\qquad\qquad
 m^2_{s_3}\approx  m_3^2\,\frac{(1+\sin2\theta_3)^2}{\cos^22\theta_3}\,.
  \label{nu_mass3}
\end{eqnarray}
The mass parameter $m_3$ can be derived from Eqs.\,(\ref{angle00}) and (\ref{msd}) as
\begin{eqnarray}
 m^2_3\simeq\frac{1}{4}\frac{\cos^22\theta_3}{\sin2\theta_3}\,(\Delta S^2_{31}-\Delta m^2_{31})\,,
  \label{m3}
\end{eqnarray}
where $\Delta S^2_{kj}\equiv m^2_{s_k}-m^2_{s_j}$.
Clearly, within the range Eq.\,(\ref{angle}) a value of $\theta_3$ going around $\pi/4$ can realize the inverted mass hierarchy (IH), $m_{\nu_2}>m_{\nu_1}\gg m_{\nu_3}$, while a value around $\pi/2$ favors the degenerate normal mass ordering (DNO), $m_{\nu_3}\gtrsim m_{\nu_2}\gtrsim m_{\nu_1}$, and degenerate inverted mass ordering (DIO), $m_{\nu_2}\gtrsim m_{\nu_1}\gtrsim m_{\nu_3}$. 
Hence, in this picture, neutrino oscillations can be described by ten parameters: six (two independent $\Delta m^2_{\rm Atm}$, $\Delta m^2_{\rm Sol}$, three mixing angles $\theta_{12},\theta_{13},\theta_{23}$, and a Dirac CP phase $\delta_{CP}$) associated with the standard three-active neutrino oscillations\,\cite{PDG} and four ($\Delta m^2_{1,2}$, $\Delta S^2_{31}, \theta_3$) responsible for the new oscillations involving sterile neutrino \,\footnote{$\Delta m^2_3$ is determined via $\Delta m^2_3=\Delta m^2_{3i}-\Delta S^2_{3i}\approx-\Delta S^2_{3i}$ for $\Delta S^2_{3i}\gg|\Delta m^2_{3i}|$ in the limit $\Delta m^2_i\rightarrow0$ with $i=1,2$.}. 

Assuming $\Delta m^2_{1(2)}\ll\Delta m^2_{\rm Sol}\ll |\Delta m^2_{\rm Atm}|\ll|\Delta m^2_{3}|$, we expect that the effects of the pseudo-Dirac neutrinos for the first and second generations can be detected through ABL oscillation experiments\,\cite{Ahn:2016hbn, Ahn:2016hhq}, whereas that for the third generation can be measured through SBL oscillation experiments (or possibly long baseline oscillation experiments).
The mass splittings will manifest themselves through very long wavelength oscillations characterized by the $\Delta m^2_{1(2)}$ as well as very short wavelength oscillations characterized by the $\Delta m^2_{3}$. 
The mass splitting $\Delta m^2_{3}$ could be limited by the active neutrino mass orderings with a requirement of $|\Delta m^2_3|\gg\Delta m^2_{\rm Atm}$:
 \begin{eqnarray}
   |\Delta m^2_3|\gg2.6\times10^{-3}\,\text{eV}^2\,,
  \label{pD_bound1}
 \end{eqnarray}
where the hierarchical mass orderings ($m_{\nu_2}>m_{\nu_1}\gg m_{\nu_3}$ and $m_{\nu_3}\gg m_{\nu_2}> m_{\nu_1}$) are used.
And since the mass splittings $\Delta m^2_{1(2)}$ can modify the large mixing angle solution of the solar neutrino oscillations, they should be limited and detailed fits imply a bound\,\cite{deGouvea:2009fp} 
 \begin{eqnarray}
   \Delta m^2_{1(2)}<1.8\times10^{-12}\,\text{eV}^2\,\,\text{at}\,3\sigma\,.
  \label{D_bound}
 \end{eqnarray}
Thus, we simply ignore $\Delta m^2_{1(2)}$ in the study of short baseline neutrino oscillations .

\section{Constraints on the new oscillation parameters}
 In this section, we present how the mixing parameters $\theta_3$ and $\Delta S^2_{31}$(or $\Delta m^2_3$, actually, the mass scale of the third generation of sterile neutrino) could be constrained through the sum of three active neutrinos $\sum m_\nu$\,\cite{Ade:2015xua, Moscibrodzka:2016ofe, Kohlinger:2017sxk}, and the effective neutrino masses in $\beta$-decay\,\cite{Kraus:2012he} as well as $0\nu\beta\beta$-decay\,\cite{Giunti:2012tn} experiments.

Oscillation experiments are unfortunately insensitive to the absolute scale of neutrino masses. Whereas cosmology is mostly sensitive to the total energy density in neutrinos, directly proportional to the sum of the active neutrino masses $\sum m_{\nu}=m_{\nu_1}+m_{\nu_2}+m_{\nu_3}$.
We will mainly focus on cosmological observations as a probe of the absolute neutrino mass scale.
Using Eq.\,(\ref{m3}), the active neutrino masses in Eq.\,(\ref{nu_mass}) can be expressed in terms of the new parameters ($\Delta S^2_{31}$, $\theta_3$) and the two known mass squared differences of oscillation experiments ($\Delta m^2_{\rm Atm}$, $\Delta m^2_{\rm Sol}$) as
\begin{eqnarray}
 m^2_{\nu_1}&=&(\Delta S^2_{31}-\Delta m^2_{\rm Atm}-\frac{1}{2}\Delta m^2_{\rm Sol})\frac{(1-\sin2\theta_3)^2}{4\sin2\theta_3}-\Delta m^2_{\rm Atm}-\frac{1}{2}\Delta m^2_{\rm Sol}\,,\nonumber\\
 m^2_{\nu_2}&=&(\Delta S^2_{31}-\Delta m^2_{\rm Atm}-\frac{1}{2}\Delta m^2_{\rm Sol})\frac{(1-\sin2\theta_3)^2}{4\sin2\theta_3}-\Delta m^2_{\rm Atm}+\frac{1}{2}\Delta m^2_{\rm Sol}\,,\nonumber\\
 m^3_{\nu_3}&=&(\Delta S^2_{31}-\Delta m^2_{\rm Atm}-\frac{1}{2}\Delta m^2_{\rm Sol})\frac{(1-\sin2\theta_3)^2}{4\sin2\theta_3}\,.
 \label{nu_mass2}
\end{eqnarray}

Cosmological and astrophysical measurements provide powerful constraints on the sum of neutrino masses complementary to those from accelerators and reactors. There are several upper limits on the sum of active neutrino masses coming from the CMB data and weak lensing data:
 \begin{eqnarray}
 0.06\,[{\rm eV}]\lesssim\sum_{i}m_{\nu_i}<\left\{
       \begin{array}{lll}
         0.340\sim0.715\,{\rm eV}\, ,& \hbox{CMB PLANCK\,\cite{Moscibrodzka:2016ofe}} \\
         0.170\,{\rm eV}\, ,& \hbox{CMB PLANCK+BAO\,\cite{Ade:2015xua}} :\\
         3.3\,{\rm eV}\, ,& \hbox{Weak lensing-only\,\cite{Kohlinger:2017sxk}}\,
       \end{array}
     \right.
  \label{nu_sum_b}
 \end{eqnarray}
where a lower limit could be provided by the neutrino oscillation measurements.

\begin{figure}[h]
\includegraphics[width=8.0cm]{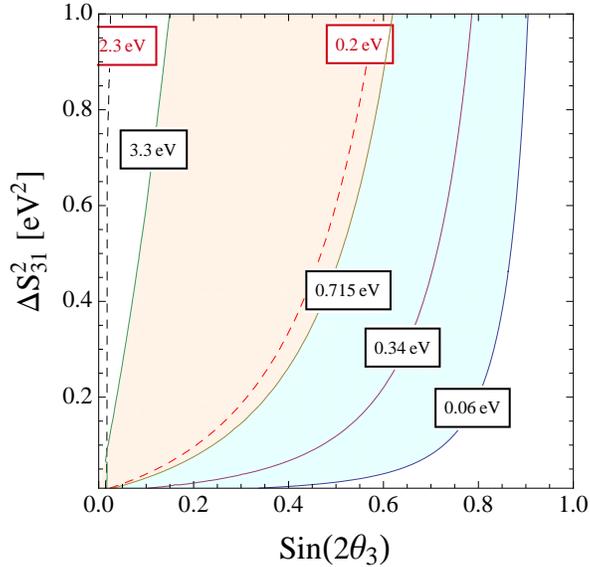}
\caption{\label{Fig1} Contour plots in  the parameter space ($\Delta S^2_{31}$, $\sin2\theta_3$) for 
fixed values of $\sum_{i=1,2,3} m_{\nu}$ (solid lines) and  $m_{\bar{\nu}_e}$ probed in tritium $\beta$ decay (dotted lines). The black-dotted line corresponds to the upper bound $m_{\bar{\nu}_e}<2.3$ eV\,\cite{Kraus:2004zw}, whereas the red-dotted line to a future sensitivity of $m_{\bar{\nu}_e}\lesssim0.20$\,\cite{Mertens:2015ila}. For $\sum_{i=1,2,3} m_{\nu}$, we take the values from
Eq.(\ref{nu_sum_b}).}
\end{figure}

In order to extract the new physics effects, first we investigate the influence of $\Delta S^2_{31}$ and $\sin2\theta_3$ on $\sum_{i}m_{\nu_i}$ by imposing the experimental results on $\Delta m^2_{\rm Atm}$, $\Delta m^2_{\rm Sol}$ and constraint  given in Eq.\,(\ref{nu_sum_b}) into Eq.\,(\ref{nu_mass2}).
Contour plots in the parameter space ($\Delta S^2_{31}$, $\sin2\theta_3$) for fixed values of
$\sum m_{\nu}$  (solid lines) and   $m_{\bar{\nu}_e}$ probed in tritium $\beta$ decay(dotted lines) are presented in Fig.\,\ref{Fig1}, where a lower limit for the sum of the neutrino masses, $\sum_{i=1}^{3} m_{\nu_i}\gtrsim0.06$ eV could be provided by the neutrino oscillation measurements; upper limits $0.715$ eV and $3.3$ eV at $95\%$ CL are given by Planck Collaboration\,\cite{Ade:2015xua} and weak lensing-only\,\cite{Kohlinger:2017sxk}, respectively, in Eq.\,(\ref{nu_sum_b}).
 In the plot\footnote{The SBL anomalies including MiniBooNE data may indicate the existence of eV-mass sterile neutrino if those are interpreted as new oscillation effects, while present cosmological data coming from CMB + large-scale structure and big bang nucleosynthesis (BBN) do not prefer extra fully thermalized sterile neutrinos in the eV-mass range since they violate the hot dark matter limit on the neutrino mass\,\cite{Archidiacono:2012ri}.
The amount of thermalisation $\Delta N^{\rm eff}_\nu$ as a
function of neutrino parameters (mass splitting, mixing, and initial lepton asymmetry) has been quantitatively derived in Ref.\,\cite{Hannestad:2012ky}, implying that the parameter space of $(\Delta m^2_3$, $\theta_3$) responsible for the existence of an eV-mass sterile neutrino is allowed by requiring such sterile neutrino does not or partially equilibrium at the BBN epoch when the initial lepton asymmetry is large\,\cite{Abazajian:2004aj, Hannestad:2012ky}.} 
we consider only eV-mass scale of sterile neutrino since too heavy neutrino is conflict with cosmology $\Delta N^{\rm eff}_\nu<0.2$ at $95\%$ CL\,\cite{Cyburt:2015mya, Hannestad:2012ky}.

The existence of sterile neutrino with the eV mass can also be constrained by $\beta$-decay experiments\,\cite{Kraus:2012he} and by $0\nu\beta\beta$ decay experiments\,\cite{Giunti:2012tn}.
The two types of mass ordering, discussed above, should be compatible with the existing constraints on the absolute scale of neutrino masses. The most sensitive experiment to  search for  the new physics effects  in $\beta$-decay is to use the tritium decay process $^3{\rm H}\rightarrow \,^3{\rm He}+e^-+\bar{\nu}_e$. Non-zero neutrino masses distort the measurable spectrum of the emitted electron. The most stringent  upper bounds on the $\bar{\nu}_e$ mass, $m_{\bar{\nu}_e}$, have been obtained from direct searches in the Mainz\,\cite{Kraus:2004zw} and Troitsk\,\cite{Aseev:2011dq} experiments at 95\% CL:
  \begin{eqnarray}
 m_{\bar{\nu}_e}=\Big(\sum_{k=1}^3|U_{ek}|^2m^2_{\nu_k}\Big)^{\frac{1}{2}}<
\left\{
\begin{array}{c}
2.30\,{\rm eV}\,(\hbox{Mainz}) \\
2.05\,{\rm eV}\,(\hbox{Troitsk})
 \end{array} 
\right.\,.
   \label{tritum}
 \end{eqnarray}
 In Fig.\,\ref{Fig1} the dotted lines show contour for
the neutrino mass in tritium $\beta$ decay $m_{\bar{\nu}_e}$ as a function of $\Delta S^2_{31}$ and $\sin2\theta_3$ with Eq.\,(\ref{angle}), where 
the black-dotted line corresponds to  the upper bound $m_{\bar{\nu}_e}<2.3$ eV\,\cite{Kraus:2004zw}, whereas the red-dotted line to a future sensitivity of $m_{\bar{\nu}_e}\lesssim0.20$\,\cite{Mertens:2015ila}.
As seen in Fig.\,\ref{Fig1}, the cosmological bounds given in Eq.\,(\ref{nu_sum_b})  are still tighter than the constraints from tritium $\beta$ decay. The upcoming KATRIN experiment\,\cite{Mertens:2015ila} planned to reach the sensitivity of $m_{\bar{\nu}_e}\sim0.20$ eV will probe the region of the quasi-degenerate mass spectrum of the active neutrinos. 

On the other hand, the $0\nu\beta\beta$-decay rate\,\cite{Aalseth:2004hb} effectively measures the absolute value of the $ee$-component of the effective
neutrino mass matrix ${\cal M}_{\nu}$ in Eq.\,(\ref{nu_matr}). In the basis where the charged lepton mass matrix is
real and diagonal, the $0\nu\beta\beta$-decay rate can be expressed as
\begin{eqnarray}
 ({\cal M}_{\nu})_{ee}=\sum_{k=1}^3W^\ast_\nu  \begin{pmatrix} m_{\nu_k}I_3 & 0_3  \\ 0_3 & m_{s_k}I_3  \end{pmatrix}W^\dag_{\nu}\Big|_{ee}\,.
\end{eqnarray}
Since the two mass eigenstates of first and second generations in each pseudo-Dirac pair have opposite $CP$ parity,  the third generation dominantly contributes to the $\beta\beta0\nu$-decay rate.
For $\big||m_{\nu_3}|-|m_{s_3}|\big|\gg\big||m_{\nu_k}|-|m_{s_k}|\big|$ with $\theta_{k}\approx0$ (for $k=1,2$), the $0\nu\beta\beta$-decay rate, $m_{\beta\beta}\equiv|({\cal M}_{\nu})_{ee}|$, is approximately given by
\begin{eqnarray}
 m_{\beta\beta}\approx\frac{1}{2}\sin^2\theta_{13}\big|(\sin2\theta_3+1)\,|m_{\nu_3}|+(\sin2\theta_3-1)\,|m_{s_3}|\big|\,.
 \label{nubb}
\end{eqnarray}
Using Eqs.\,(\ref{nu_mass3}) and (\ref{m3}) one can easily see that $\beta\beta0\nu$-decay rate
becomes almost zero, $m_{\beta\beta}\approx0$.
Hence if the $\beta\beta0\nu$-decay rate is measured in near future the model would explicitly be excluded\,\footnote{Note that the claim of observation of $0\nu\beta\beta$-decay of $^{76}_{32}$Ge\,\cite{KlapdorKleingrothaus:2004wj} is strongly disfavored by the
recent results of the GERDA experiment\,\cite{Agostini:2013mzu}.}.
  
\section{short baseline neutrino anomalies}
Now, let us study how our model can help to resolve the so-called short baseline neutrino anomalies in terms of neutrino oscillations.
To see how the new sterile neutrino states cause such new oscillations at short-baselines with neutrino trajectory less than $1<$ km,
let us bring out a conversion probability of new oscillations with the help of the neutrino mixing matrix Eq.\,(\ref{Wnu}). The conversion probability\,\footnote{The transition probability of $\nu_\alpha\rightarrow \nu_{s_i}$ between sterile and active neutrinos due to oscillations of active flavor $\nu_\alpha$ (with $\alpha=e,\mu,\tau$) with sterile neutrinos $\nu_{s_i}$ (with $i=1,2,3$), see Eq.\,(\ref{Wnu}), is given by $P_{\nu_\alpha\rightarrow \nu_{s_i}}\simeq\sum^2_{k=1}U^\ast_{i k}U_{\alpha k}\tilde{U}^\ast_{\alpha k}\tilde{U}_{i k}\sin^2(\frac{\Delta m^2_k}{4\pi}L)+U^\ast_{i 3}U_{\alpha 3}\tilde{U}^\ast_{\alpha 3}\tilde{U}_{i 3}\cos^22\theta_3\sin^2(\frac{\Delta m^2_3}{4\pi}L)$ where $\tilde{U}\equiv U_R$ in Eq.\,(\ref{Wnu}), $\Delta m^2_{kj}\simeq m^2_{\nu_k}-m^2_{s_j}$ and $\Delta S^2_{kj}\simeq m^2_{s_k}-m^2_{\nu_j}$ with $k>j=1,2,3$ are used. For non-vanishing $\tilde{U}_{\mu3}$, the new mass squared difference $\Delta m^2_3$ could be constrained by the probability of $\nu_\mu\rightarrow \nu_{s_i}$, while for tiny or vanishing $\tilde{U}_{\mu3}$ one could not constrain the $\Delta m^2_3$. In addition, the mass squared difference $\Delta m^2_3$ could be constrained by the disappearance  of muon-type neutrinos and antineutrinos produced in the atmosphere, see Eq.\,(\ref{ter01}) and Fig.\,\ref{Fig6}, where the mixing $\theta_3$ deviated from the maximal mixing $\pi/4$ is involved.}  between the massive neutrinos that a neutrino eigenstate $\nu_a$ becomes eigenstate $\nu_b$ follows from the time evolution of mass eigenstates as
\begin{eqnarray}
P_{\nu_a\rightarrow\nu_b}(W_\nu, L, E)&=&\Big|(W^\ast_\nu\,e^{-i\frac{\hat{\cal M}^2_\nu}{2E}L}W^T_\nu)_{ab}\Big|^2\,,
\label{proba}
\end{eqnarray}
where $a,b=e,\mu,\tau,s_1,s_2,s_3$, $L$ is the distance between the neutrino detector and the neutrino
source, $E$ is the neutrino energy, and $\hat{\cal M}_\nu\equiv W^T_\nu{\cal M}_\nu W_\nu$.
We are interested in the flavor conversion between the active neutrinos $\nu_e,\,\nu_\mu,\,\nu_\tau$ satisfying the condition of Eq.\,(\ref{D_bound}) which leads to $\theta_{1(2)}\approx0$. From Eq.\,(\ref{proba}) the flavor conversion probability between the three-active neutrinos can explicitly be expressed in terms of the oscillation parameters $\theta$, $\Delta m^2$, $L$, $E$, and mixing components $U_{\alpha i}$ of the $3\times3$ PMNS matrix as
\begin{eqnarray}
P_{\nu_\alpha\rightarrow\nu_\beta}&=&\delta_{\alpha\beta}-\sum^2_{k=1}|U_{\alpha k}|^2|U_{\beta k}|^2\sin^2\Big(\frac{\Delta m^2_k}{4E}L\Big)-|U_{\alpha 3}|^2|U_{\beta 3}|^2\sin^2\Big(\frac{\Delta m^2_3}{4E}L\Big)\,\cos^22\theta_3\nonumber\\
&-&\sum_{k>j}{\rm Re}\big[U^\ast_{\beta k}U_{\beta j}U^\ast_{\alpha j}U_{\alpha k}\big]\Big[(1+\delta_{k3}\sin2\theta_3)\Big\{\sin^2\Big(\frac{\Delta m^2_{kj}}{4E}L\Big)+\sin^2\Big(\frac{\Delta Q^2_{kj}}{4E}L\Big)\Big\}\nonumber\\
&&\qquad\qquad\qquad\qquad+(1-\delta_{k3}\sin2\theta_3)\Big\{\sin^2\Big(\frac{\Delta S^2_{kj}}{4E}L\Big)+\sin^2\Big(\frac{\Delta Q^2_{jk}}{4E}L\Big)\Big\}\Big]\nonumber\\
&+&\frac{1}{2}\sum_{k>j}{\rm Im}\big[U^\ast_{\beta k}U_{\beta j}U^\ast_{\alpha j}U_{\alpha k}\big]\Big[(1+\delta_{k3}\sin2\theta_3)\Big\{\sin\Big(\frac{\Delta m^2_{kj}}{2E}L\Big)+\sin\Big(\frac{\Delta Q^2_{kj}}{2E}L\Big)\Big\}\nonumber\\
&&\qquad\qquad\qquad\qquad+(1-\delta_{k3}\sin2\theta_3)\Big\{\sin\Big(\frac{\Delta S^2_{kj}}{2E}L\Big)-\sin\Big(\frac{\Delta Q^2_{jk}}{2E}L\Big)\Big\}\Big]\label{osc01}\,,
\end{eqnarray}
where $\Delta Q^2_{kj}\equiv m^2_{\nu_k}-m^2_{s_j}$, and $\delta_{k3}=1$ for $k=3$ and $0$ for $k\neq3$. In the model the mixing parameters $\theta$ and $\Delta m^2$ are determined by nature, so experiments should choose $L$ and $E$ to be sensitive to oscillations through a given $\Delta m^2$.
As expected, in the limit of $m_{s_i}\rightarrow m_{\nu_i}$ and $\theta_{i}\rightarrow0$ ($i=1,2,3$), 
$P_{\nu_\alpha\rightarrow\nu_\beta}$ becomes the standard form of conversion probability for three active neutrinos in vacuum, as shown in Ref.\,\cite{PDG}.
The model has interesting features listed below under the assumption of CPT invariance:
\begin{itemize}
\item{
From Eq.\,(\ref{osc01}), we see that
 $P_{\nu_\alpha\rightarrow\nu_\alpha}+P_{\nu_\alpha\rightarrow\nu_\beta}+P_{\nu_\omega\rightarrow\nu_\alpha}\leq1$ with $\alpha\neq\beta\neq\omega=e,\mu,\tau$, whereas the probabilities for the standard three-active neutrino oscillations satisfy the relation $ P_{\nu_\alpha\rightarrow\nu_\alpha}+P_{\nu_\alpha\rightarrow\nu_\beta}+P_{\nu_\omega\rightarrow\nu_\alpha}=1$.}
\item{The new oscillation effects attributed to $\Delta m^2_i$, $\Delta S^2_{3k}$ and $\Delta Q^2_{k3}$ with $i=1,2,3$ and $k=1,2$ in Eq.\,(\ref{osc01}) can be 
 maximum in the limit of  $\theta_3\rightarrow\pi/2$ (favored by the DNO and DIO), which can be relevant to short baseline neutrino experiments.}
\item{ In the limit of  $\theta_3\rightarrow\pi/4$ (favored by the IH, see Eq.\,(\ref{nu_mass2})), the oscillatory term involving $\Delta m^2_{k}$ (with $k=1,2$) in Eq.\,(\ref{osc01}) can give new oscillation effects only applicable to ABL oscillations. And thus, it is expected that this case could not provide a solution to the SBL anomalies.}
\end{itemize}

To investigate the effects of new oscillations due to the new sterile neutrinos, we present approximated forms of neutrino oscillation probability based on the formula given by Eq.\,(\ref{osc01}), which are relevant to interpreting short baseline neutrino anomalies.
At a distance satisfying  $L\ll 4\pi E/\Delta m^2_{\rm Sol}, 4\pi E/\Delta m^2_{1(2)}$, 
the survival probability for $\bar{\nu}_e$ is approximated as
\begin{eqnarray}
P_{\bar{\nu}_e\rightarrow\bar{\nu}_e}&\approx&1-\sin^4\theta_{13}\sin^2\Big(\frac{\Delta m^2_{3}}{4E_{\bar{\nu}_e}}L\Big)\cos^22\theta_3\nonumber\\
&-&\frac{1}{2}
\sin^22\theta_{13}\Big[(1-\sin2\theta_3)\sin^2\Big(\frac{\Delta S^2_{31}}{4E_{\bar{\nu}_e}}L\Big)+(1+\sin2\theta_3)\sin^2\Big(\frac{\Delta m^2_{31}}{4E_{\bar{\nu}_e}}L\Big)\Big].
\label{reac}
\end{eqnarray}
We note that new oscillatory terms vanish in the limit of $\theta_3=\pi/4$, whereas they can reach maximum in the limit of $\theta_3=\pi/2$.
In the limit that $\frac{\Delta m^2_{31}}{4 E_{\nu}}L$ is negligible, but   $\frac{4 E_{\nu}}{\Delta S^2_{31}} \simeq -\frac{4 E_{\nu}}{\Delta m^2_{3}} \sim L $, the ${\nu}_e$ disappearance probability $P_{\bar{\nu}_e\rightarrow\bar{\nu}_e}$ becomes
\begin{eqnarray}
P_{\bar{\nu}_e\rightarrow\bar{\nu}_e}\approx 1-\Big[\sin^4\theta_{13}\cos^22\theta_3+\frac{1}{2}
\sin^22\theta_{13}(1-\sin2\theta_3)\Big]\sin^2\Big(\frac{\Delta S^2_{31}}{4E_{\bar{\nu}_e}}L\Big)\,.
\label{reac}
\end{eqnarray}
Similarly, the probabilities $P_{\bar{\nu}_{\mu}\rightarrow\bar{\nu}_e}$ and $P_{\bar{\nu}_{\mu}\rightarrow\bar{\nu}_{\mu}}$ are approximately given by
\begin{eqnarray}
P_{\bar{\nu}_\mu\rightarrow\bar{\nu}_e} &\approx& \frac{1}{4}\sin^22\theta_{13}\sin^2\theta_{23}\big[2-2\sin2\theta_3-\cos^22\theta_3\big]\sin^2\Big(\frac{\Delta S^2_{31}}{4E_{\bar{\nu}_e}}L\Big)\,,\label{lsnd}\\
P_{\bar{\nu}_\mu\rightarrow\bar{\nu}_\mu}&\approx& 1-\sin^2\theta_{23}\cos^2\theta_{13}\Big[\sin^2\theta_{23}\cos^2\theta_{13}(\cos^22\theta_3-2\sin2\theta_3+2)\nonumber\\   
&& \qquad\qquad\qquad\qquad\qquad-2(1-\sin2\theta_3)\Big]\sin^2\Big(\frac{\Delta S^2_{31}}{4E}L\Big)\,,
\label{lsnd2}
\end{eqnarray}
The above formulae for the probabilities can be applied to not only  the experimental data from LSND and MiniBooNE but also the reactor neutrino flux anomaly.
The expressions Eqs.\,(\ref{reac},\ref{lsnd},\ref{lsnd2}) can be compared with those in the 3+1 model\,\cite{white, Gariazzo:2015rra} given by
\begin{eqnarray}
P_{\bar{\nu}_{\alpha}\rightarrow\bar{\nu}_{\beta}}^{3+1}\simeq 
\sin^2 2\theta_{\alpha \beta} \sin^2\left( \frac{\Delta m^2_{41}L}{4E}\right)\,,\qquad 
P_{\bar{\nu}_{\alpha}\rightarrow\bar{\nu}_{\alpha}}^{3+1}\simeq 
1-\sin^2 2\theta_{\alpha \alpha} \sin^2\left( \frac{\Delta m^2_{41}L}{4E}\right)\,,
\label{3+1osc}
\end{eqnarray}
where $\alpha, \beta = e, \mu, \tau, s$.
The oscillation amplitudes depend only on the absolute values of the elements in the forth column of the mixing matrix in the $3+1$ model,
\begin{eqnarray}
\sin^2 2 \theta_{\alpha \beta} = 4|U_{\alpha 4}|^2|U_{\beta 4}|^2, ~(\alpha \neq \beta)\,,\qquad
\sin^2 2 \theta_{\alpha \alpha} = 4|U_{\alpha 4}|^2(1-|U_{\alpha 4}|^2)\,.
\end{eqnarray}
Then, $\Delta S^2_{31}$ plays the same role of $\Delta m^2_{41}$ in (\ref{3+1osc}), and
$\sin^2 2 \theta_{\alpha \beta}$ and $\sin^2 2 \theta_{\alpha \alpha}$ correspond to the parameters multiplied in front of oscillatory terms in Eqs.\,(\ref{reac},\ref{lsnd},\ref{lsnd2}).

On the other hand, for the baselines optimized by the oscillation parameters $\Delta m^2_{31}\sim2.5\times10^{-3}\,{\rm eV}^2$ and $E_{\bar{\nu}_e}\sim$ MeV, {\it i.e.} $4\pi E/\Delta S^2_{31}\ll L\ll 4\pi E/\Delta m^2_{\rm Sol}, 4\pi E/\Delta m^2_{1(2)}$, the  antineutrino probability in Eq.\,(\ref{reac}) is approximately given by 
\begin{eqnarray}
P_{\bar{\nu}_e\rightarrow\bar{\nu}_e}\approx1-\frac{1}{2}\sin^4\theta_{13}\cos^22\theta_3
-\frac{1}{2}
\sin^22\theta_{13}\Big[\frac{1-\sin2\theta_3}{2}+(1+\sin2\theta_3)\sin^2\Big(\frac{\Delta m^2_{31}}{4E_{\bar{\nu}_e}}L\Big)\Big]\,.
\label{reacNew}
\end{eqnarray}
where the terms including $\Delta S^2_{31}$ are averaged out.

Now, let us examine how the SBL anomalies can be resolved or alleviated by the formulae given above.
Note that all plots in what follows are based on the exact formulae in Eq.\,(\ref{osc01}) and for comparison we use the best-fit values of NO in Eq.\,(\ref{mixing_angle}) unless otherwise noted.
\subsection{New Interpretation of Reactor Neutrino Results}
\begin{figure}[h]
\begin{minipage}[h]{7.3cm}
\epsfig{figure=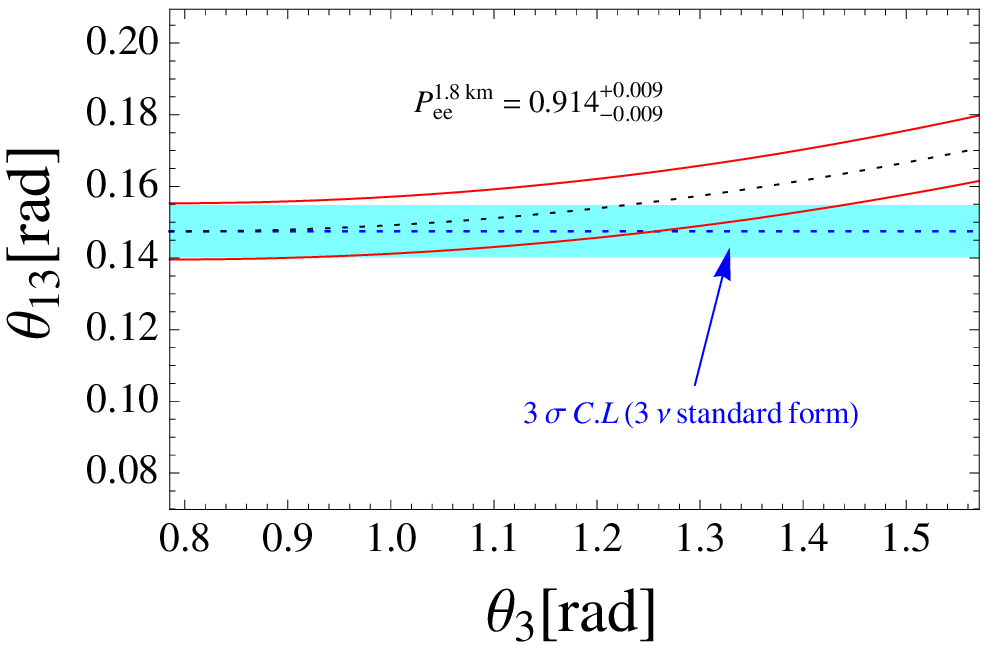,width=8.5cm,angle=0}
\end{minipage}
\hspace*{1.0cm}
\begin{minipage}[h]{7.3cm}
\epsfig{figure=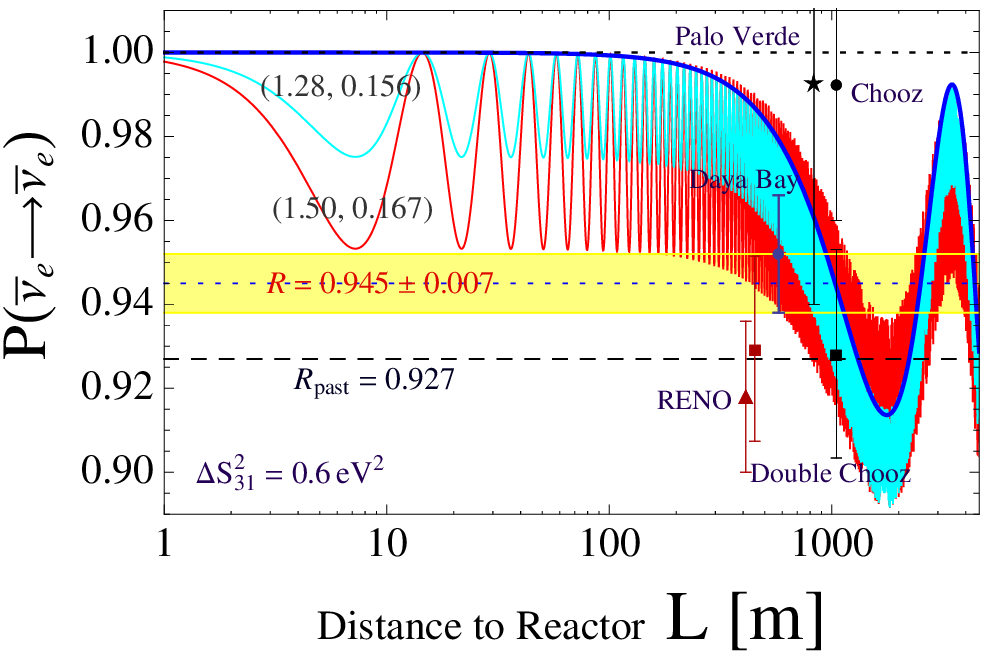,width=8.5cm,angle=0}
\end{minipage}
\caption{\label{Fig2} Left plot represents $\theta_{13}$ {\it vs.} $\sin(2\theta_3)$, where the cyan band(horizontal dotted-line) stands for $\theta_{13}[^\circ]=8.45^{+0.45}_{-0.45}$ at $3\sigma$ (best-fit value)given  in Eq.\,(\ref{mixing_angle}). Right plot represents  $P_{\bar{\nu}_e\rightarrow\bar{\nu}_e}$ {\it vs.} $L$ [meters] for $E_{\bar{\nu}_e}=3.5$ MeV where  the standard form of $P_{3\nu}(\bar{\nu}_e\rightarrow\bar{\nu}_e)$ for a fixed $\theta_{13}=8.45^\circ$ is plotted by the blue-solid curve, and the red and cyan-sine curves are obtained from the exact $P_{\bar{\nu}_e\rightarrow\bar{\nu}_e}$ for $(\theta_3, \theta_{13})=(1.50, 0.167)$ and ($1.28, 0.156$) with $\Delta S^2_{31}=0.6\,{\rm eV}^2$. Error bars represent recent several measurements of $R$, where the yellow band stands for the most recent world average\,\cite{Adey:2018qct} and the black-dashed horizontal line for the previous world average\,\cite{white}.}
\end{figure}

In order to probe the effects of the new sterile neutrino from the experimental results obtained at the reactor neutrino experiments, Daya Bay, RENO and Double Chooze (see Ref.\,\cite{Nu2018}), we interpret them  in terms of neutrino oscillations including the new sterile neutrino by using the new reactor antineutrino probability in Eq.\,(\ref{reacNew}).
Adopting $\Delta m^2_{31}$ by the values determined from atmospheric neutrino oscillation, we can obtain new values of $\theta_{13}$ along with $\theta_3$ 
by equating Eq.\,(\ref{reacNew}) with  the value of $\bar{\nu}_e$ survival probability for the three-active neutrino oscillation given by
 \begin{eqnarray}
 P_{3\nu}(\bar{\nu}_e\rightarrow \bar{\nu}_e)\approx1-\sin^22\theta_{13}\sin^2\Big(\frac{\Delta m^2_{31}}{4E_{\bar{\nu}_e}}L\Big)=0.914^{+0.009}_{-0.009}\,,
 \label{reac0}
\end{eqnarray}
where the numerical result is obtained by taking $L=1.8\,{\rm km}, E_{\bar{\nu}_e}=3.5\,{\rm MeV}, \Delta m^{2}_{31}=2.5\times10^{-3}\,{\rm eV}^2$, and at $3\sigma$ $\theta_{13}[^\circ]=8.45^{+0.45}_{-0.45}$ given in Eq.\,(\ref{mixing_angle}).
Then, $\theta_{13}$ becomes a function of $\theta_3$. In the right panel of Fig.\,\ref{Fig2}, each depth of sinusoidal curves is proportional to $\sin^22\theta_{13}$ as indicated in Eq.\,(\ref{reacNew}). 
The left panel of Fig.\,\ref{Fig2} shows the behavior of $\theta_{13}$ driven by Eq.\,(\ref{reacNew}) as a function of $\theta_3$, where the values of $\theta_{13}$ are enhanced by around $6\%$ at $\theta_3=1.28$ and $13\%$ at $\theta_3=1.50$ (recalling that the value of $\theta_3\lesssim1.28$ for $\Delta S^2_{31}=0.6\,{\rm eV}^2$ is constrained by cosmological data\footnote{If the sum of active neutrino masses constrained by PLANCK data in Eq.\,(\ref{nu_sum_b}) is relaxed making the value of $\theta_3$ large {\it i.e.} $\theta_3>1.28$, the value of $\theta_{13}$ can also be enhanced and in turn the LSND/MiniBooNE anomaly could be explained as seen in the following section.}, see Fig.\,\ref{Fig1}):
\begin{eqnarray}
 \theta_{13}[^\circ]= 8.96^{+0.45}_{-0.49}~~\text{at}~\theta_3=1.28\,[{\rm rad}];\qquad\theta_{13}[^\circ]=9.54^{+0.49}_{-0.53}~~\text{at}~\theta_3=1.50\,[{\rm rad}]\,.
\label{nTheta13}
\end{eqnarray}
In the right panel of Fig.\,\ref{Fig2}, we  fix $E_{\bar{\nu}_e}=3.5$ MeV and $\Delta S^2_{31}=0.6\,{\rm eV}^2$ and the cyan and red plots correspond to the cases of  $(\theta_3, \theta_{13})=(1.28, 0.156)$  and $(1.50, 0.167)$ , respectively. 
Note that the value of $\theta_{13}$ corresponds to the central value evaluated at the given value of $\theta_3$ with the new oscillation formula Eq.\,(\ref{reacNew}), and the red-triangle, red-square (black-square), blue-circle, black-star, and black-circle error bars represent the values of the parameter $R$ defined by the ratio of reactor antineutrino flux to the theoretical prediction obtained from
the RENO \,\cite{Nu2018}, Double Chooz\,\cite{Abe:2014bwa, Nu2018}, Daya Bay\,\cite{Adey:2018qct}, Palo Verde\,\cite{Boehm:2001ik}, and Chooz\,\cite{Apollonio:2002gd}, where the error bars represent the experimental uncertainties. The values of $R$ from RENO, Double Chooz, Daya Bay, Palo Verde, and Chooz have been obtained by subtracting the effects of $\theta_{13}$-driven oscillations. And the yellow band stands for the world average of $R$ updated after including Daya Bay result $R=0.945\pm0.007 ({\rm exp.})$\,\cite{Adey:2018qct}, compared with the past global average $R_{\rm past}=0.927$\,\cite{white} indicated as black-dashed horizontal line, where the uncertainty of common reactor model is $\pm 0.023$ with respect to the Huber-Mueller model.
\subsection{Reactor Neutrino Flux Anomaly}
The reactor antineutrino anomaly\,\cite{Mention:2011rk} is the experimental result presenting a deficit of the rate of $\bar{\nu}_e$ in several SBL reactor neutrino experiments with $L\sim(10-100)$ m 
and $E_{\bar{\nu}_e}\sim$ MeV.
In reactor neutrino experiments, electron antineutrinos are detected through the inverse neutron decay process $\bar{\nu}_e+p\rightarrow n+e^+$ in liquid-scintillator detectors. To interpret the deficit of observed reactor neutrino fluxes relative to the prediction (Huber-Mueller model\,\cite{Mueller:2011nm}) in terms of neutrino oscillation including the new sterile neutrinos, it is relevant to use the probability given by Eq.\,(\ref{reac}).
 We note that the reactor neutrino flux anomaly
is not clearly explained at $L\lesssim500$ m as in the 3+1 model\,\cite{white, Gariazzo:2015rra}: 
Eq.\,(\ref{reac}) clearly depends on the electron antineutrino energy $E_{\bar{\nu}_e}$ and its flight length $L$, and the nature of sterile neutrino associated with $\Delta S^2_{31}$ and $\theta_3$. Once experimental inputs $L$ and $E_{\bar{\nu}_e}$ are fixed the allowed regions of $\Delta S^2_{31}$ and $\theta_3$ can be obtained from $P_{\bar{\nu}_e\rightarrow\bar{\nu}_e}$ in Eq.\,(\ref{reac}), constraints by $\sum m_\nu$ in Eq.\,(\ref{nu_sum_b}), and $m_{\bar{\nu}_e}$ in Eq.\,(\ref{tritum}) as shown in Fig.\,\ref{Fig1}. Note that the values of the model parameters $\theta_3$ and $\Delta S^2_{31}$ favored by the Planck data in Eq.\,(\ref{nu_sum_b}) are not conflict with the NEOS ($L=24$ m) and DANSS ($L=10.7\rightarrow12.7$ m) results\,\cite{Ko:2016owz}\,\footnote{The recent NEOS and DANSS results\,\cite{Ko:2016owz} have a tension with the Gallium and reactor anomalies in the 3+1 model\,\cite{white}, while their results show a preference on our model predictions.}.

\subsection{LSND anomaly and MiniBooNE data}
\label{lsnd000}
\begin{figure}[h]
\begin{minipage}[h]{7.3cm}
\epsfig{figure=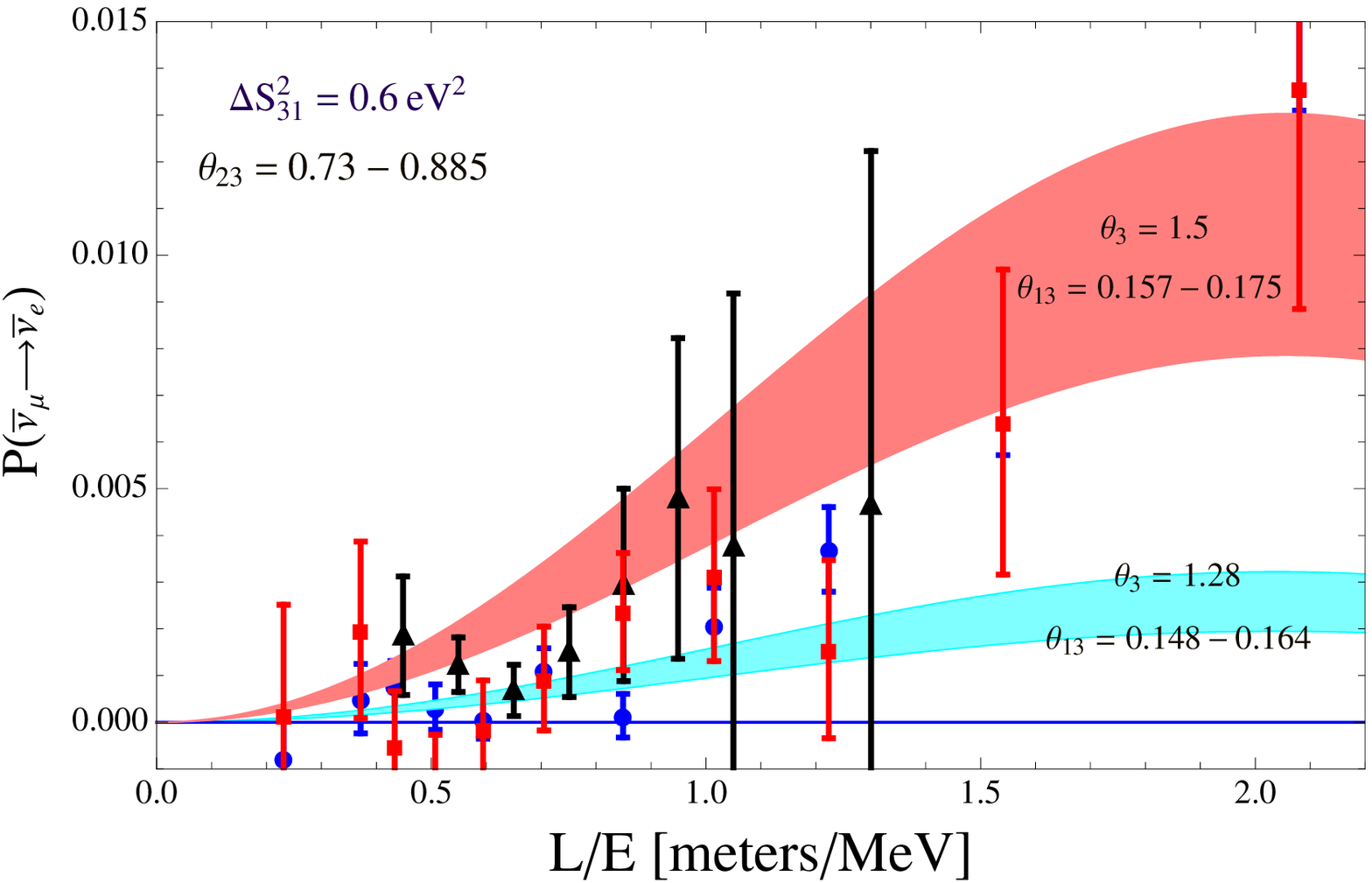,width=8.5cm,angle=0}
\end{minipage}
\hspace*{1.0cm}
\begin{minipage}[h]{7.3cm}
\epsfig{figure=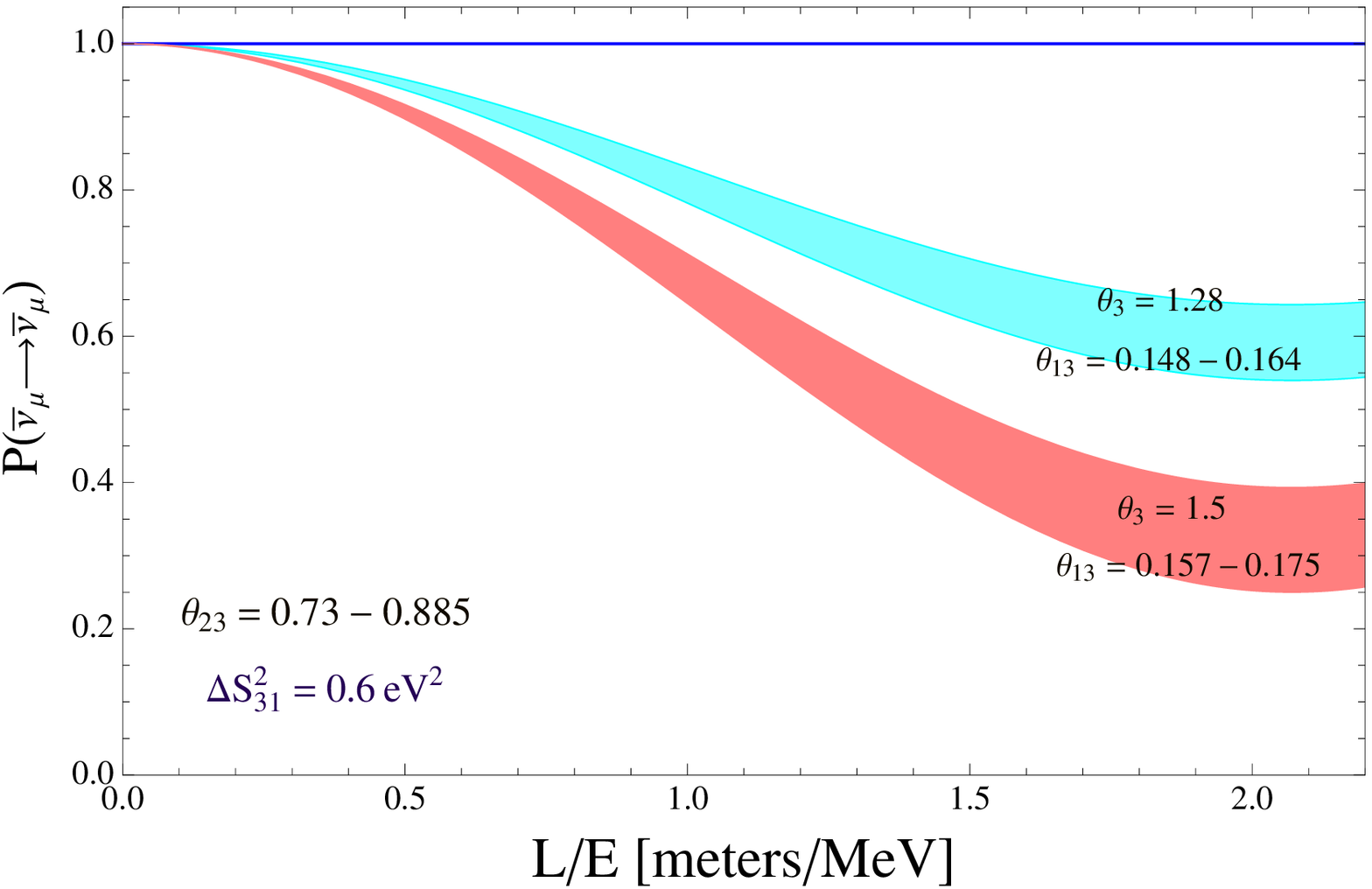,width=8.5cm,angle=0}
\end{minipage}
\caption{\label{Fig3} Plots of $P_{\bar{\nu}_\mu\rightarrow\bar{\nu}_e}$ (left) and $P_{\bar{\nu}_\mu\rightarrow\bar{\nu}_\mu}$ (right) {\it vs.} $L/E$ [m/MeV]. In the left panel 
data points with blue(red)-bar  correspond to neutrino (antineutrino) mode in  MiniBooNE\,\cite{Aguilar-Arevalo:2018gpe}  and those with black-bar to  the LSND\,\cite{Athanassopoulos:1995iw} data.  The horizontal-blue lines stand for the standard form of probability for $\bar{\nu}_{e}$ appearance (left) and $\bar{\nu}_{\mu}$ disappearance (right), 
while the red- and cyan-band curves represent the new conversion probabilities in Eq.\,(\ref{osc01}) for the sum of active neutrino mass $\sum m_\nu=2.673$ and $0.705$ eV, respectively. The band width is due to the uncertainty of $\theta_{23}=41.8\sim50.7^\circ$ and the new ranges of $\theta_{13}$ in Eq.\,(\ref{nTheta13}).}
\end{figure}
The LSND experiment\,\cite{Athanassopoulos:1995iw} reported observation of  a statistically suggestive excess of $\bar{\nu}_e$ events in a beam of $\bar{\nu}_\mu$ produced by $\mu^+$ decay at rest, $\mu^+\rightarrow e^++\nu_e+\bar{\nu}_\mu$ with $3.8\sigma$ significance\,\footnote{The similar KARMEN
 collaboration\,\cite{Armbruster:2002mp} did not measure any excess of $\bar{\nu}_e$ events over the background at a mean distance $L\simeq17.7$ m with $E_{\bar{\nu}_\mu}=12\sim52.8$ MeV, which did not fully exclude the LSND result.}.
The MiniBooNE experiments also observed  $\nu_e$ and $\bar{\nu}_e$ appearance in $\nu_\mu$ and $\bar{\nu}_\mu$ beams, respectively\,
having the same $L/E$ as in LSND\,\cite{Aguilar-Arevalo:2018gpe}. 
In the left panel of  Fig.\,\ref{Fig3}, the black error bars stand for the LSND\,\cite{Athanassopoulos:1995iw} data and the blue (red) error bars stand for the MiniBooNE data excess in (anti-)neutrino mode, at the baseline 541 m from the beryllium target, in particular, in the interval of energies $200<E_\nu<475$ MeV, which corresponds to $L/E$ range beyond that probed in the LSND experiment\,\cite{Aguilar-Arevalo:2018gpe}. It has been shown that  the observed excesses in MiniBooNE experiment are in agreement with the LSND result, and provide a good fit to a large $\Delta m^2$ solution in a two-neutrino oscillation framework, even though the two experiments have completely different neutrino energies, neutrino fluxes, reconstruction, backgrounds, and systematic uncertainties.

The LSND and MiniBooNE excesses could be explained by $\bar{\nu}_\mu\rightarrow\bar{\nu}_e$ oscillation whose probability is given by Eq.\,(\ref{lsnd}).
As in the case of $3+1$ model\,\cite{white, Gariazzo:2015rra},  to explain the excess of $\bar{\nu}_e$ events, we need $\Delta S^2_{31} \sim O(1) {\rm eV}^2$ and large values of $\theta_3 (> \pi/4)$. In particular, when  $\theta_3\rightarrow\pi/2$ favored by the degenerate case (DNO and DIO),
$P_{\bar{\nu}_\mu\rightarrow\bar{\nu}_e}\simeq\frac{1}{4}\sin^22\theta_{13}\sin^2\theta_{23}\sin^2\Big(\frac{\Delta S^2_{31}}{4E_{\bar{\nu}_e}}L\Big)  \lesssim 0.012$ for the LSND and MinibooNE experiments. But this case leads to a large value of the sum of three active neutrino masses, as shown in Fig.\,\ref{Fig3}. The left panel of Fig.\,\ref{Fig3} presents plots of $P_{\bar{\nu}_\mu\rightarrow\bar{\nu}_e}$ versus $L/E ({\rm m/MeV})$ for two bench mark points of $\theta_3=1.5$ and $1.28$ for a given $\Delta S^2_{31}=0.6 {\rm eV}^2$. In the numerical estimation, we vary the parameter space $(\theta_{13},\, \theta_{23})$ in the ranges of $(0.157-0.175,\, 0.730-0.885)$ and $(0.148-0.164,\,0.730-0.885)$  which correspond to  $\sum_{i=e,\mu,\tau} m_{\nu_i}=2.673$ and $0.705$ eV, respectively, for two bench mark points.
%
%
It is worthwhile to note that the parameter space of ($\Delta S^2_{31}$, $\theta_3$), as presented in Fig.\,\ref{Fig1}, is constrained through $\sum_{i=e,\mu,\tau} m_{\nu_i}$ by the cosmological data  and the effective neutrino mass in tritium $\beta$-decay.
 However, statistical uncertainties have to be reduced by gaining more data in order to confirm our model in the following two respects: (i) 
 the two data points in the left panel of Fig.\,\ref{Fig3} from MiniBooNE at $L/E\gtrsim1.5$ (m/MeV) seem to favor $\sum m_\nu\gtrsim0.705$ eV, which is disfavored by Planck Collaboration (TT+lowP) at $95\%$ CL\,\cite{Moscibrodzka:2016ofe} while still not excluded by the weak lensing only data\,\cite{Kohlinger:2017sxk}, (ii)  it seems that the $\nu_e$ (blue-bars) and $\bar{\nu}_e$ (red-bars) modes in MiniBooNE data, in principle, could be discriminated by considering the CP violating term in Eq.\,(\ref{osc01}). In the model setup, however, it seems not possible to discriminate between them within the short baseline due to the CP violating terms proportional to $-4\sin(\Delta m^2_{21}L/2E)$.

In addition, searching for the $\bar{\nu}_e$ appearance in LSND and MiniBooNE also implies $\bar{\nu}_\mu$ disappearance whose oscillation probability is given by Eq.\,(\ref{lsnd2}). In the right panel of Fig.\,\ref{Fig3}, we present how the survival probability $P_{\bar{\nu}_\mu\rightarrow\bar{\nu}_\mu}$ behaves along with $L/E({\rm m/MeV})$ for the same cases in the left panel of Fig.\,\ref{Fig3}. The large deviation from $P_{\bar{\nu}_\mu\rightarrow\bar{\nu}_\mu}=1$ at large $L/E$ as shown in the right panel of Fig.\,\ref{Fig3} is the characteristic feature of this model we consider, which makes our model different from $3+1$ model.
%
Accelerator based experiments for $\bar{\nu}_\mu$ (or $\nu_\mu$) disappearance, such as MINOS and
MINOS+ neutrino experiments\,\cite{minos}, may be sensitive to oscillations involving sterile neutrinos\,\cite{Sousa:2015bxa} for the regions of $10^{-2}\lesssim L/E ({\rm km/GeV})\lesssim0.75$ at near detector ($L=1.04$ km) and $8\lesssim L/E({\rm km/GeV})\lesssim7\times10^2$ at far detector ($L=735$ km). The former region can cover $L/E$ values of order 1m/MeV in the right panel of Fig.\,\ref{Fig3}, while the latter can cover range of $L/E\sim{\cal O}(10\sim100)$ [km/GeV], comparing with the results of the atmospheric muon neutrino events observed\,\footnote{The recent results from the MINOS and MINOS+ far detector data in Neutrino 2018\,\cite{Nu2018} including the result of IceCube DeepCore\,\cite{Aartsen:2014yll} seem to be in agreement with the expected in the three neutrino standard form in the range $20\lesssim L/E[{\rm km/GeV}]\lesssim2000$, while in their past results\,\cite{Sousa:2015bxa, Whitehead:2016xud} there are some large deficit data points which seem to be in agreement with the oscillating signatures of light sterile neutrinos, see the cyan-curve in the left panel of Fig.\,\ref{Fig6}.} in Super-Kamiokande\,\cite{Ashie:2004mr} and the left plot in Fig.\,\ref{Fig6}. 
To see such an oscillation dip due to new sterile neutrino as shown in the right panel of Fig.\,\ref{Fig3}, we need
 the neutrino baseline $\sim6$ km (14 km) for detector having $\sim3$ GeV (7 GeV) peak energy with
$\Delta S^2_{31}=0.6 \,{\rm eV}^2$.

In addition, such $\nu_\mu$ disappearance oscillation effect could be observed at $L/E\sim1.2$ [m/MeV] with $E_{\nu_\mu}\sim0.5$ GeV and $L=600$ m in the Short-Baseline Neutrino (SBN) Program experiment\,\cite{Bass:2017qhe} for the same model parameters in Fig.\,\ref{Fig3}.
And, at the far detector ($L=735$ km) a possibility on searching for signatures of sterile neutrinos in the $\nu_\mu$ disappearance by using Eq.\,(\ref{lsnd2}) will be considered in section\,\ref{nu_mu}.

\subsection{Earth matter effects}
Let us explore the Earth matter effect\,\cite{Wolfenstein:1977ue, Barger:1980tf} by examining the propagation of atmospheric neutrinos produced in cosmic-ray air showers from the Earth's atmosphere to the inside of the Earth. When muon neutrinos pass through the Earth matter,  the MSW effect\,\cite{Wolfenstein:1977ue} should be taken into account. Path-length ranges from $10$ km to $1.27\times10^4$ km depending on arrival zenith angle. The matter density encountered by neutrinos propagating is on average $\rho_\oplus\sim3\,g/cm^3$ in the Earth's crust and outer mantle, $\sim5\,g/cm^3$ in the inner mantle, and between 10 and $\sim13\,g/cm^3$ in the core\,\cite{Dziewonski:1981xy}. Muon neutrino oscillations modified due to matter effects can produce distinctive signatures of sterile neutrinos in the large
set of high energy atmospheric neutrino data (both in the TeV energy window from IceCube\,\cite{TheIceCube:2016oqi} and at lower energy from DeepCore\,\cite{Aartsen:2014yll}). 
For $E_{\nu}>100$ GeV, three-active neutrino oscillation length  larger than the diameter of the Earth and can be neglected.
In order to find appropriate physics parameters ($\theta, \Delta m^2, L, E$) for atmospheric neutrino oscillations, we consider the effective Hamiltonian in-matter ${\cal H}^m$ in flavor basis, which has the form of $6\times6$ matrix
\begin{eqnarray}
 {\cal H}^m&=&\frac{1}{2E_m}\left[W^\ast_\nu{\left(\begin{array}{cc}
 m^{2}_{\nu_k}I_3 & 0_3  \\
 0_3 &  m^{2}_{s_k}I_3
 \end{array}\right)} W^T_\nu+{\left(\begin{array}{cc}
 A_{\alpha}I_3 & 0_3  \\
 0_3 &  0_{3}
 \end{array}\right)}\right]\,,
\label{mat}
\end{eqnarray}
where $k=1,2,3$ and $\alpha=e, \mu, \tau$. Here the parameters $A_\alpha=2E_mV_\alpha$ is a measure of the importance of matter effect with the matter-induced effective potential; $V_e$, $V_\mu$, $V_\tau$, and $V_s=0$ are the potentials experienced by the electron, muon, tau, and sterile neutrinos respectively, and $E_m$ is the neutrino energy in matter. For anti-neutrinos $V_\alpha\rightarrow-V_\alpha$. $\nu_e$'s have charged-current (CC) interactions with electrons and neutral-current (NC) interactions with electrons and nucleons, $V_e=\sqrt{2}G_F(N_e-N_n/2)$, while $\nu_\mu$'s and $\nu_\tau$'s have only NC interactions, $V_\mu=V_\tau=\sqrt{2}G_F(-N_n/2)$, and any $\nu_s$'s have no interactions, $V_s=0$, where Fermi's constant, $G_F$, and the average electron and neutron densities along the neutrino path, $N_e$ and $N_n$, respectively.

The mass matrix of the massive neutrinos in matter can be diagonalized through a new unitary mixing matrix $W_m$ ,
\begin{eqnarray}
 W_m={\left(\begin{array}{cc}
 U_L & 0_3  \\
 0_3 &  U_R 
 \end{array}\right)}{\left(\begin{array}{cc}
 V_1 & iV_1  \\
 V_2 &  -iV_2 
 \end{array}\right)}{\left(\begin{array}{cc}
 e^{i\phi_k}\cos\theta^m_kI_3 & -e^{i\phi_k}\sin\theta^m_kI_3  \\
 e^{-i\phi_k}\sin\theta^m_kI_3 &  e^{-i\phi_k}\cos\theta^m_kI_3
 \end{array}\right)}\,.
\label{mat1}
\end{eqnarray}
The diagonalization of ${\cal H}^m$ by the unitary matrix $W_m$ in matter gives a condition 
\begin{eqnarray}
 A_\alpha|U_{\alpha k}|^2=\Delta m^2_k\frac{\sin2(\theta^m_k-\theta_k)}{\cos2\theta^m_{k}}\,,
\label{mat2}
\end{eqnarray}
and its effective mass-squared eigenvalues in matter which are positive
\begin{eqnarray}
 \tilde{m}^2_{\nu_k}&=(m^2_{\nu_k}+m^2_{s_k})\cos^2(\theta^m_k-\theta_k)-m^2_{s_k}\cos2(\theta^m_k-\theta_k)+\frac{\Delta m^2_k}{2}\frac{\sin2(\theta^m_k-\theta_k)}{\cos2\theta^m_k}(1+\sin2\theta^m_k)\nonumber\\
 \tilde{m}^2_{s_k}&=(m^2_{\nu_k}+m^2_{s_k})\cos^2(\theta^m_k-\theta_k)-m^2_{\nu_k}\cos2(\theta^m_k-\theta_k)+\frac{\Delta m^2_k}{2}\frac{\sin2(\theta^m_k-\theta_k)}{\cos2\theta^m_k}(1-\sin2\theta^m_k)\,.
\label{mat3}
\end{eqnarray}
\begin{figure}[h]
\includegraphics[width=10.0cm]{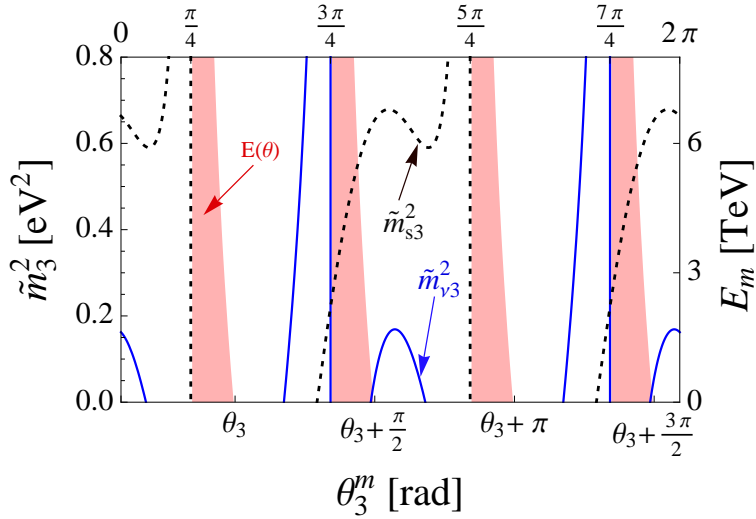}
\caption{\label{Fig40} Plots of the effective mass-squared $\tilde{m}^2_{s_3}$ (black-dotted curves) and $\tilde{m}^2_{\nu_3}$ (blue-solid curves) in Eq.\,(\ref{mat3}) and the effective energy $E(\theta)\equiv E_m(\theta^m_3)$ in Eq.\,(\ref{mat4})  as a function of $\theta^m_3$. The shaded regions stand for the allowed parameter space by the matter effect, while the white regions for vacuum-like oscillations.}
\end{figure}
For $\theta_{1(2)}\sim0$ corresponding to $\Delta m^2_{1(2)}<10^{-12}\,{\rm eV}^2$ in Eq.\,(\ref{D_bound}) we assume no Earth matter effects occurs, leading to $\theta^m_k\rightarrow\theta_k$.
Then from Eq.\,(\ref{mat2}) the energy of upward-going atmospheric muon neutrinos in matter can be derived as
\begin{eqnarray}
E_{m}\simeq\frac{1.6\,{\rm TeV}}{\sin^2\theta_{23}\cos^2\theta_{13}}\left(\frac{\Delta m^2_{3}}{-0.6\,{\rm eV}^2}\right)\Big(\frac{5\,g/cm^3}{\rho_\oplus}\Big)\frac{\sin2(\theta^m_3-\theta_3)}{\cos2\theta^m_3}\,,
\label{mat4}
\end{eqnarray}
and $\sin2(\theta^m_3-\theta_3)\rightarrow\sin2(\theta_3-\theta^m_3)$ for $\bar{\nu}_\mu$, indicating that the energy $E_m$  varies with the matter mixing angle $\theta^m_3$. 
Note that as $\theta^m_3\rightarrow\theta_3+n\pi/2$ (with $n=0,1,2,3,...$), the matter effect gets faded.
In Fig.\,\ref{Fig40} we plot  the effective mass-squared $\tilde{m}^2_{s_3}$ (black-dotted lines), $\tilde{m}^2_{\nu_3}$  (blue-solid lines) and $E_m$  (red-bands) in terms of $\theta^m_3$.
Thanks to the positiveness of  $\tilde{m}^2_{s_3}$, $\tilde{m}^2_{\nu_3}$ and $E_m$,
they are predicted for limited regions of $\theta_3^m$ as can be seen in Fig.\,\ref{Fig40}.
The lines overlapped with the red bands represent he effective masses affected by Earth matter effect,
%
In the limit of $E_m\rightarrow0$, $\tilde{m}^2_{\nu_3}$ approached to $m^2_{\nu_3}$, implying that  matter effect  becomes negligible for the muon neutrinos with energies $\gg1$ GeV passing through the interior of Earth. Hence, it is expected that the $\nu_\mu$( or $\bar{\nu}_\mu)$ disappearance probability for high energy ($\gg1$ GeV) muon neutrinos passing through the Earth interior to search for the oscillation signatures of light sterile neutrinos is the same as the one in vacuum-like derived in Eq.\,(\ref{proba}).

Recently, the high-energy IceCube detector  has measured the atmospheric muon neutrino spectrum 
at energy  $E_{\nu_\mu}=320\,{\rm GeV}\sim20$ TeV
in  hope of finding the oscillation signatures of light sterile neutrinos\,\cite{TheIceCube:2016oqi}, but no evidence for anomalous $\nu_\mu$ or $\bar{\nu}_\mu$ disappearance is observed. The results coincide with the model prediction on the matter effects at the same energy window, contrary to the models in Refs.\,\cite{Dentler:2018sju, Asaadi:2017bhx, Kostelecky:2003cr, Gninenko:2009ks, Liao:2016reh, Carena:2017qhd} which has a tension with $\nu_\mu$ (or $\bar{\nu}_\mu$) disappearance experiments. And  the IceCube sub-detector DeepCore at energy window of atmospheric muon neutrinos $10\sim100$ GeV\,\cite{Aartsen:2014yll} can also have potential to search for the signature of light sterile neutrinos as discussed in the next section.

\section{new effects in solar and atmospheric oscillations}
Now, let us examine the oscillation effects  due to the sterile neutrinos  on the long baseline experiments
such as KamLAND\,\cite{PDG}, T2K\,\cite{Abe:2014nuo}, MINOS and MINOS+ Collaboration\,\cite{minos}, solar neutrino and atmospheric neutrino experiments\,\cite{Aartsen:2014yll, Ashie:2004mr}.
%

\subsection{ New effects in $\nu_e$ disappearance from KamLAND, T2K and Solar neutrino oscillation}
For the long baseline such as KamLAND experiment\,\cite{kamland}, the survival probability of $\bar{\nu}_e$ events in the model we consider is approximately given by 
\begin{eqnarray}
P_{\bar{\nu}_e\rightarrow\bar{\nu}_e}&\approx&1-\frac{1}{2}(\sin^22\theta_{13}
-\sin^4\theta_{13}\cos^22\theta_{3})
-\cos^4\theta_{13}\sin^22\theta_{12}\sin^2\Big(\frac{\Delta m^2_{21}}{4E}L\Big)
\label{long03}
\end{eqnarray}
where we assume CP invariance and the term $\sin^2\big(\frac{\Delta m^2_3}{4E}L\big)$  is averaged out for  long baseline  (e.g. $\langle L\rangle\simeq180$ km).
%
%
Applying Eq.\,(\ref{long03}) to KamLAND data\,\cite{Abe:2008aa}, we can extract the value of $\theta_3$ with precise measurements of $\Delta m^2_{21}$ and $\theta_{12}$. However, we see that Eq.\,(\ref{long03}) is almost the same as the expression for the standard oscillation probability for three-active neutrinos because the new term concerned with $\theta_3$  is negligible due to the tiny value of $\sin^4\theta_{13}\simeq5\times10^{-4}$.
Thus, new effects due to the sterile neutrinos on KamLAND, T2K experiments and solar neutrino oscillation
are negligible.

Since the T2K experiment at the ND280 near detectors covers $L/E$ values of order $1\,{\rm m}/{\rm MeV}$, an information such as $\theta_3$ in Eq.\,(\ref{long03}) on sterile neutrinos can also be extracted. It has performed a search for $\nu_e$ disappearance in a neutrino beam whose $\nu_e$ component is peaked at an energy of 500 MeV\,\cite{Abe:2014nuo}. From Eq.\,(\ref{long03}) in the limit $L\ll4\pi E_{\nu_e}/\Delta m^2_{21}$ the $\nu_e$ disappearance probability driven by the sterile neutrino can  easily be obtained for the baseline of 280 m, and it shows that its effect on sterile neutrino is negligible due to the tiny value of $\sin^4\theta_{13}$.

\subsection{New effects in $\nu_\mu$ disappearance from SuserK and IceCube}
\label{nu_mu}
\begin{figure}[h]
\begin{minipage}[h]{7.3cm}
\epsfig{figure=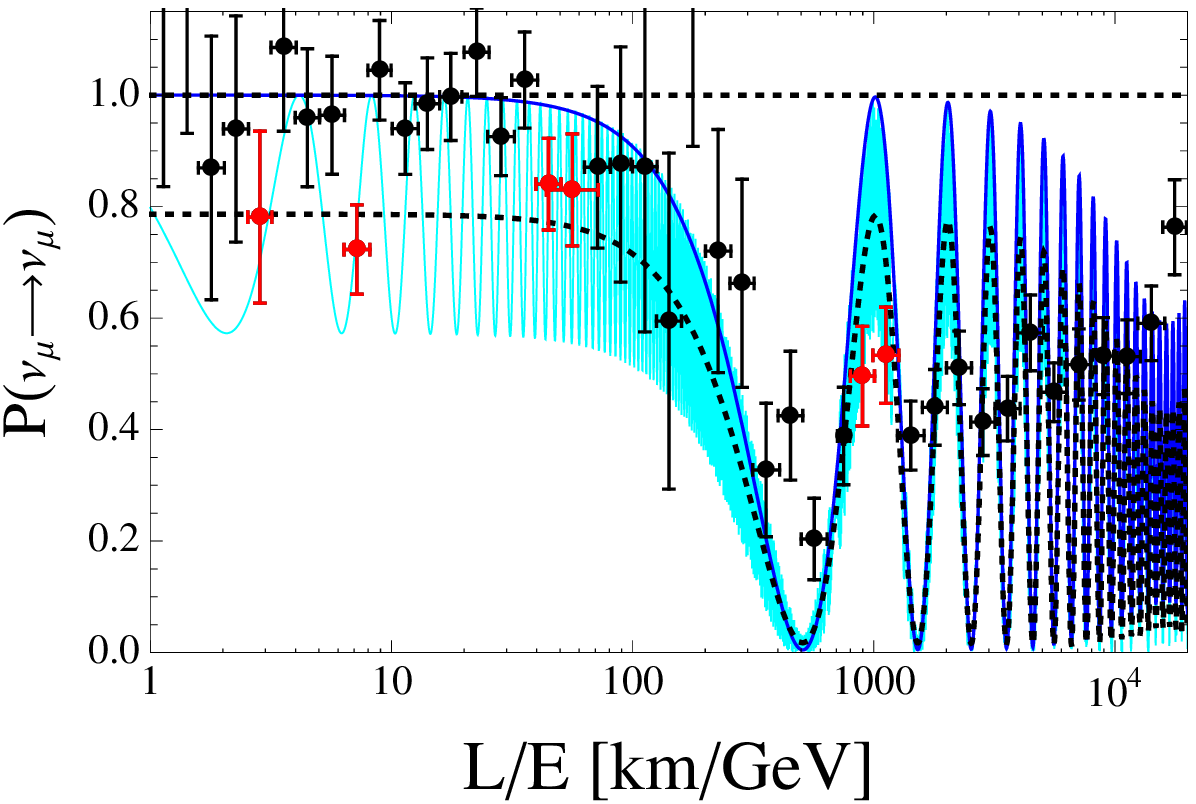,width=8.5cm,angle=0}
\end{minipage}
\hspace*{1.0cm}
\begin{minipage}[h]{7.3cm}
\epsfig{figure=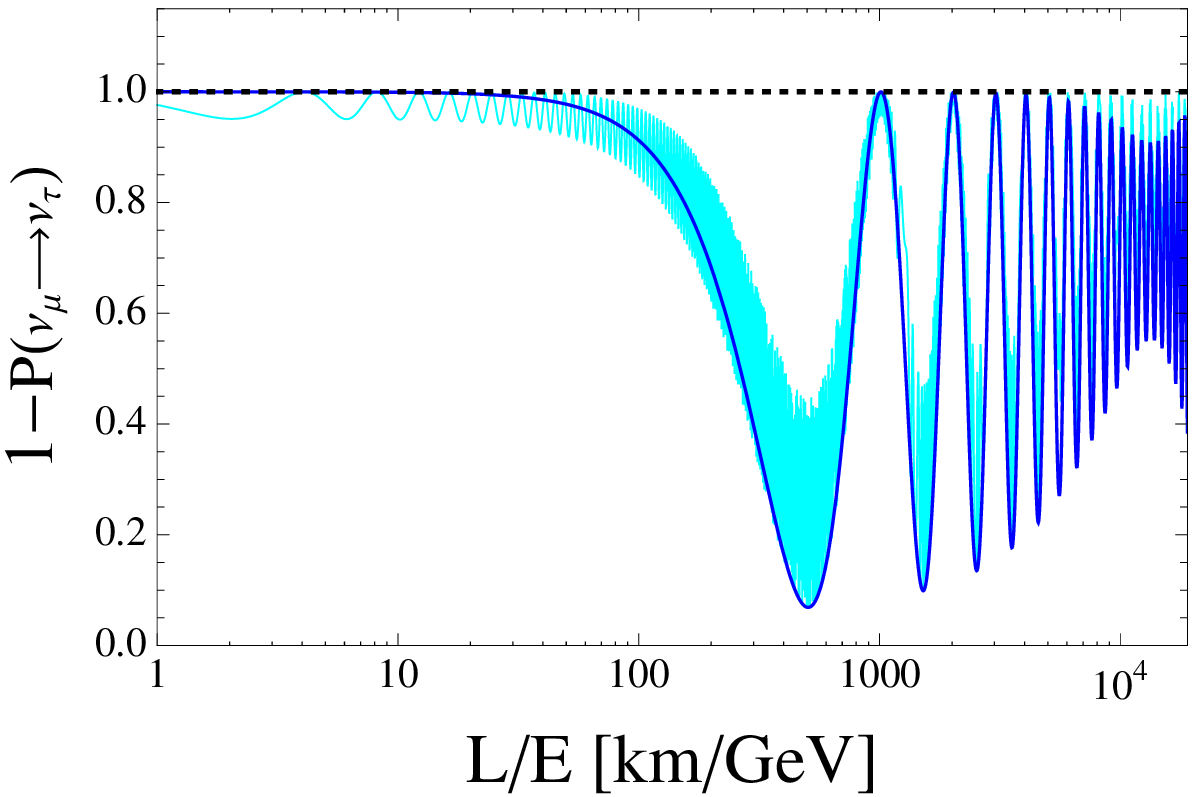,width=8.5cm,angle=0}
\end{minipage}
\caption{\label{Fig6} Plots of $P_{\bar{\nu}_\mu\rightarrow\bar{\nu}_\mu}$ (left) and $1-P_{\nu_\mu\rightarrow\nu_\tau}$  (right) {\it vs.} $L/E$ [km/GeV], where the blue sinusoidal curves stand for the standard forms of the three-active neutrino probabilities and the black-dotted curve for $\sin^2\Big(\frac{\Delta S^2_{3_i}}{4E}L\Big)$ being averaged out.  In the left plot the data points represent the atmospheric muon neutrino events observed in Super-Kamiokande\,\cite{Ashie:2004mr}.}
\end{figure}

In addition to the TeV muon neutrinos discussed in previous section, the IceCube Collaboration can also have potential to search for the signature of light sterile neutrinos by observing atmospheric neutrinos in the tens-of-GeV range through its sub-detector DeepCore (for reference, see Ref.\,\cite{Aartsen:2014yll}).
The atmosphere of the Earth is constantly being bombarded by cosmic rays, primarily made up of protons and helium nuclei produced by atmospheric objects\,\cite{PDG}. These cosmic rays have been observed over a wide range of energy, from 1 GeV to $10^{11}$ GeV. We recall atmospheric neutrinos with energy of $\sim$ a few GeV which are mostly produced by primary cosmic rays with energy of $\sim100$ GeV\,\cite{PDG}. 
According to the three-active neutrino oscillation probability, large deficits of muon neutrino have been observed in upward-going events and at $L/E\sim230$ km/GeV\,\cite{PDG} as shown by blue sinusoidal curves in  Fig.\,\ref{Fig6} which is well consistent with the Super-Kamiokande's atmospheric neutrino data\,\cite{Ashie:2004mr}.
For the baselines $L\ll 4\pi E/\Delta m^2_{\rm Sol}, 4\pi E/\Delta m^2_{1(2)}$, which is relevant to the atmospheric neutrino, the survival and conversion probability of muon neutrino are approximately given by
\begin{eqnarray}
P_{\nu_\mu\rightarrow\nu_\mu}&\approx&1-\sin^4\theta_{23}\cos^4\theta_{13}\cos^22\theta_3 \sin^2\Big(\frac{\Delta m^2_3}{4E}L\Big)
-2\sin^2\theta_{23}\cos^2\theta_{13}(1-\sin^2\theta_{23}\cos^2\theta_{13})\nonumber \\
&&\qquad \times \Big[(1+\sin2\theta_3)\sin^2\Big(\frac{\Delta m^2_{31}}{4E}L\Big)+(1-\sin2\theta_3)\sin^2\Big(\frac{\Delta S^2_{31}}{4E}L\Big)\Big]\,,
\label{ter01} \\
P_{\nu_\mu\rightarrow\nu_\tau}&\approx&\frac{1}{2}\sin^22\theta_{23}\cos^4\theta_{13}\Big[-\frac{1}{2}\cos^22\theta_3\sin^2\Big(\frac{\Delta m^2_3}{4E}L\Big)\nonumber\\
&&\qquad+(1+\sin2\theta_3)\sin^2\Big(\frac{\Delta m^2_{31}}{4E}L\Big)+(1-\sin2\theta_3)\sin^2\Big(\frac{\Delta S^2_{31}}{4E}L\Big)\Big]\,,
\label{ter03}
\end{eqnarray}
where $\Delta m^2_{3j}\approx\Delta Q^2_{3j}$ and $\Delta S^2_{3j}\approx|\Delta Q^2_{j3}|$ with $j=1,2$ are used.
Since $L\gg4\pi E/\Delta S^2_{31}$, $\sin^2\Big(\frac{\Delta S^2_{31}(\Delta m^2_3)}{4E}L\Big)$ is averaged out. 
As expected, the new oscillation probability for $\nu_{\mu}$ (or $\bar{\nu}_\mu)$ disappearance given in Eq.\,(\ref{ter01}) can be sizably deviated depending on the the parameters $\Delta S^2_{31}$, $\theta_3$ from the three-active neutrino oscillation probability.
From Eqs.\,(\ref{ter01}) and (\ref{ter03}), we easily see that $1-P_{\nu_\mu\rightarrow\nu_\tau}-P_{\nu_\mu\rightarrow\nu_\mu}\geq P_{\nu_e\rightarrow\nu_\mu}$ in vacuum.
At a distance $L\ll 4\pi E/\Delta m^2_{\rm Sol}, 4\pi E/\Delta m^2_{1(2)}$, 
the probability $P_{\bar{\nu}_\mu\rightarrow\bar{\nu}_e}(=P_{\nu_e\rightarrow\nu_\mu})$ is given by
\begin{eqnarray}
&&P_{\nu_e\rightarrow\nu_\mu}\approx\frac{1}{2}\sin^2\theta_{23}\sin^22\theta_{13}\Big[-\frac{1}{2}\cos^22\theta_3\sin^2\Big(\frac{\Delta m^2_3}{4E}L\Big)\nonumber\\
&&\qquad\qquad\qquad+(1+\sin2\theta_3)\sin^2\Big(\frac{\Delta m^2_{31}}{4E}L\Big)+(1-\sin2\theta_3)\sin^2\Big(\frac{\Delta S^2_{31}}{4E}L\Big)\Big]\,.
\label{ter02}
\end{eqnarray}
For $L\gg4\pi E/\Delta S^2_{31}$,  $\sin^2\Big(\frac{\Delta S^2_{31}}{4E}L\Big)$ in Eq.(\ref{ter02}) is averaged out.

The left panel of Fig.\,\ref{Fig6} shows $\nu_{\mu}$ survival probability as a function of $L/E ({\rm km/GeV})$, where the blue sinusoidal curve and the cyan rapid oscillations stand for the probability predicted from the standard form of probabilty for three active neutrinos and that from the new oscillation probability with parameters $\theta_3=1.28$ and $\Delta S^2_{31}=0.6\,{\rm eV}^2$, and the data points are atmospheric
neutrino events observed in Super-Kamiokande\,\cite{Ashie:2004mr}.
As shown in the plot the atmospheric neutrino data from Super-Kamiokande \,\cite{Ashie:2004mr} are well consistent with the new $\nu_\mu\leftrightarrow\nu_\tau$ oscillation (cyan curve), showing the first oscillation dip appeared at $\sim500$ km/GeV, and interestingly enough, the error bars with red-circle showing large deficits compared with the results in the three-active neutrino framework are also well consistent with the new oscillation curve, which may be due to the existence of new sterile neutrino.

Since the high energy muon neutrinos ($10\sim100$ GeV) passing through the interior of the Earth are little
affected by matter  in this model, it is expected that 
the IceCube sub-detector DeepCore results\,\cite{Aartsen:2014yll} are similar to that provided by Super-Kamiokande atmospheric neutrino data\,\cite{Ashie:2004mr} representing no evidence for light sterile neutrinos.
In addition to the IceCube DeepCore results\,\cite{Aartsen:2014yll}, 
the track-like muon events selected in the MINOS and MINOS+ data \,\cite{Nu2018}
shows the first oscillation dip at $\sim500$ km/GeV mainly due to the muon neutrino oscillation into not sterile neutrinos but tau neutrinos.
%
%

However, there exists a possibility to probe the existence of sterile neutrinos by comparing the results
of $P_{\nu_\mu\rightarrow\nu_\mu}$ with those of $1-P_{\nu_\mu\rightarrow\nu_\tau}$.
In fact, both results are almost equivalent for the standard oscillation for the three-active neutrinos.
But, they are different for the new oscillation affected by the sterile neutrinos.
In the right panel of Fig.\,\ref{Fig6}, we plot $1-P_{\nu_\mu\rightarrow\nu_\tau}$, where the blue sinusoidal curve and the cyan oscillation curve correspond to the standard three-active neutrino oscillations and new oscillation with $\theta_3=1.28, \Delta S^2_{31}=0.6\,{\rm eV}^2$, respectively. 
In the $\nu_\mu\leftrightarrow\nu_\tau$ oscillation, the oscillation parameters are fixed to the best-fit values\,\footnote{As the uncertainty of $\theta_{23}$ is large its impact on the result is not negligible.} for NO given in Eq.\,(\ref{mixing_angle}). 
%
Such discrepancy between $P_{\nu_\mu\rightarrow\nu_\mu}$ and $1-P_{\nu_\mu\rightarrow\nu_\tau}$would be proved through both the $\nu_\mu\rightarrow\nu_\tau$ appearance experiments and the $\nu_\mu$ disappearance experiments in the future.
Recently, the OPERA collaboration has confirmed that muon neutrinos primarily oscillate into tau neutrinos\,\cite{Agafonova:2018auq}.
In the case of $\tilde{m}^2_{\nu_3}\rightarrow m^2_{\nu_3}$, as shown in Fig.\,\ref{Fig40}, an MSW resonance is expected to occur in multi-GeV, similar to the standard oscillation for the three-active neutrino\,\cite{PDG}, for the atmospheric and accelerator $\nu_{e,\mu}$ neutrinos traveling in earth matter with the propagation eigenstates of active neutrinos\,\cite{Wolfenstein:1977ue}.

\section{Implications of IceCube data.}
The astrophysical neutrinos with very high energy fly galactic and extra galactic distances  far beyond the earth-sun distance can give us an opportunity to probe pseudo-Dirac neutrinos with very tiny mass splittings as mentioned before.
Taking into account astronomical-scale baseline satisfying $4\pi E/\Delta m^{2}_{\rm Sol, Atm}\ll L\sim4\pi E/\Delta m^2_k$ with $k=1,2$ to uncover the oscillation effects of very tiny mass splitting $\Delta m^2_k$, the probability of neutrino flavor conversion  from Eq.\,(\ref{osc01}) reads
\begin{eqnarray}
 P_{\nu_{\alpha}\rightarrow\nu_{\beta}}&=&\delta_{\alpha\beta}-\sum^2_{k=1}|U_{\alpha k}|^2|U_{\beta k}|^2\sin^2\left(\frac{\Delta m^2_{k}L}{4E}\right)\nonumber\\
 &-&\frac{1}{2}|U_{\alpha3}|^2|U_{\beta3}|^2\cos^22\theta_3-2\sum_{k>j}{\rm Re}[U^\ast_{\beta k}U_{\beta j}U^\ast_{\alpha j}U_{\alpha k}]\,,
\label{pro_ast}
\end{eqnarray}
where the oscillatory terms involving the atmospheric and solar mass-squared differences and the large mass-squared differences $\Delta m^2_{3}$, $\Delta S^2_{3k}$ and $\Delta Q^2_{3k}$ with $k=1,2$ are averaged out over such long distances..
As shown in \cite{Lunardini:2000swa},  the matter effects inside the Gamma Ray Burst (GRB) sources  as well as the earth are not significant, which makes us to consider vacuum oscillation only for astrophysical neutrinos.
%
Neutrino telescope such as IceCube\,\cite{ice-cube} observes neutrinos from extragalactic sources 
located far away from the earth and with
neutrino energy $10^5\,{\rm GeV}\lesssim E\lesssim10^7$ GeV.
Given neutrino trajectory $L$ and energy $E$, the oscillation effects  become prominent
when  $\Delta m^2_k\sim E/4\pi L$, 
where $L\equiv L(z)$ is a distance-measure with redshift $z$, which is different from comoving or luminosity distance,
given by
\begin{eqnarray}
 L(z)\equiv D_H\int^z_0\frac{dz'}{(1+z')^2\sqrt{\Omega_m(1+z')^3+\Omega_\Lambda}}\,,
 \label{}
\end{eqnarray}
where the Hubble length $D_H=c/H_0\simeq4.42$ Gpc with the results of the Planck Collaboration\,\cite{Ade:2015xua}: 
 \begin{eqnarray}
 \Omega_\Lambda=0.6911\pm0.0062\,,\quad\Omega_m=0.3089\pm0.0062\,,\quad  H_0=67.74\pm0.46\,{\rm km}\,{\rm s}^{-1}{\rm Mpc}^{-1}\,,
 \label{cosmo_const0}
 \end{eqnarray}
in which $\Omega_\Lambda$, $\Omega_m$, and $H_0$ stand for the dark energy density of the Universe, the matter density of the Universe, and the present Hubble expansion rate, respectively.
The asymptotic value of $L(z)$ is about $2.1$ Gpc achieved by large value of $z$, which means that  the smallest $\Delta m^2_k$ that can
be probed with astrophysical neutrinos with $E$ is $10^{-17}\,\mbox{eV}^2 \,(E/\rm PeV)$\,\cite{redshift}.
If this is the case, in order to observe the oscillation effects the oscillation lengths should not be  much larger than the flight length
before arriving at neutrino telescopes in earth for given tiny mass splittings, that is,
\begin{eqnarray}
 L^k_{\rm osc}\simeq\left(\frac{5\times10^{-15}\,{\rm eV}^2}{\Delta m^2_{k}}\right)\left(\frac{E}{5\times10^5{\rm GeV}}\right)8\,\text{Mpc}\lesssim8\,\text{Mpc}
\label{osc_length}
\end{eqnarray}
which means that astrophysical neutrinos with  $L\simeq8$ Mpc (flight length) and energy $E\simeq0.5\,{\rm PeV}$ would be useful
to probe the pseudo-Dirac property of neutrinos with the very tiny mass splitting $\Delta m^2_{k}\simeq5\times10^{-15}\,{\rm eV}^2$.
From Eq.\,(\ref{osc_length}), we see that given the tiny mass splittings $\Delta m^2_k=10^{-14\sim  -16}{\rm eV}^2$ with the energies around 100 TeV--1 PeV,
 a new oscillation curve at neutrino trajectory $\lesssim{\cal O}(10)$ Mpc is naively expected to occur.

On the other hand, oscillation effects induced by tiny mass splittings $\Delta m^2_k$ for the pseudo-Dirac neutrinos can affect the track-to-shower ratio for the number of shower $N_S$ and track events $N_T$ measured from IceCube experiment, which is given by \cite{Palladino:2015zua,Ahn:2016hhq},
%
%
\begin{eqnarray}
 \frac{N_T}{N_S}&=& \frac{p_T\,\big\{a_\mu\,\tilde{F}_\mu-\sum_{k=1}^2a^\mu_{k}\,\tilde{F}^\mu_{k}\big\}}{a_e\,\tilde{F}_e+a_\mu\,(1-p_T)\,\tilde{F}_\mu+a_\tau\,\tilde{F}_\tau-\sum_{k=1}^2\big\{a^e_k\,\tilde{F}^e_k+a^\mu_k\,(1-p_T)\,\tilde{F}^\mu_k+a^\tau_k\,\tilde{F}^\tau_k\big\}}\,,
 \label{ntns1}
\end{eqnarray}
where
\begin{eqnarray}
 &\tilde{F}_\alpha=\sum_{\beta}\Big\{\delta_{\alpha\beta}-\frac{1}{2}|U_{\alpha3}|^2|U_{\beta3}|^2\cos^22\theta_3-2\sum_{k>j}{\rm Re}[U^\ast_{\beta k}U_{\beta j}U^\ast_{\alpha j}U_{\alpha k}]\Big\}\,\phi^0_{\beta}\,,\nonumber\\
  &a_\alpha=4\pi\int dE\,E^{-\omega}A_\alpha(E)\,,\nonumber\\
 &\tilde{F}^\alpha_k=\sum_{\beta}|U_{\alpha k}|^2|U_{\beta k}|^2\,\phi^0_{\beta}\,,\nonumber\\
 & a^\alpha_k=4\pi\int dE\sin^2\left(\frac{\Delta m^2_{k}L}{4E}\right)E^{-\omega}A_\alpha(E)\,, \label{ak}
\end{eqnarray}
with a spectral index $\omega$,  the detector effective areas $A_{\alpha}(E)$ and initial flavor composition $\phi^0_{\beta}$.
Here $p_T$ is the probability that an observed event produced by a muon neutrino is a track event, which is mildly dependent on energy and approximately equals to $0.8$\,\cite{Aartsen:2013jdh}.
The prediction for the ratio $N_T/N_S$ depends on the initial flavor composition $\phi^0_e:\phi^0_\mu:\phi^0_\tau$ at the source which are relevant for the interpretation of observational data.
There are four well-known production mechanisms for high energy neutrinos from which the flavor compositions are given: (i) $(\frac{1}{3}: \frac{2}{3}: 0)$ for $\pi$ decay, (ii) $(\frac{1}{2}: \frac{1}{2}: 0)$ for charmed mesons decay, (iii) $(1: 0 : 0)$ for $\beta$ decay of neutrons, and (iv) $(0 : 1 : 0)$ for $\pi$ decay with damped muons.
\begin{figure}[t]
\begin{minipage}[h]{7.5cm}
\epsfig{figure=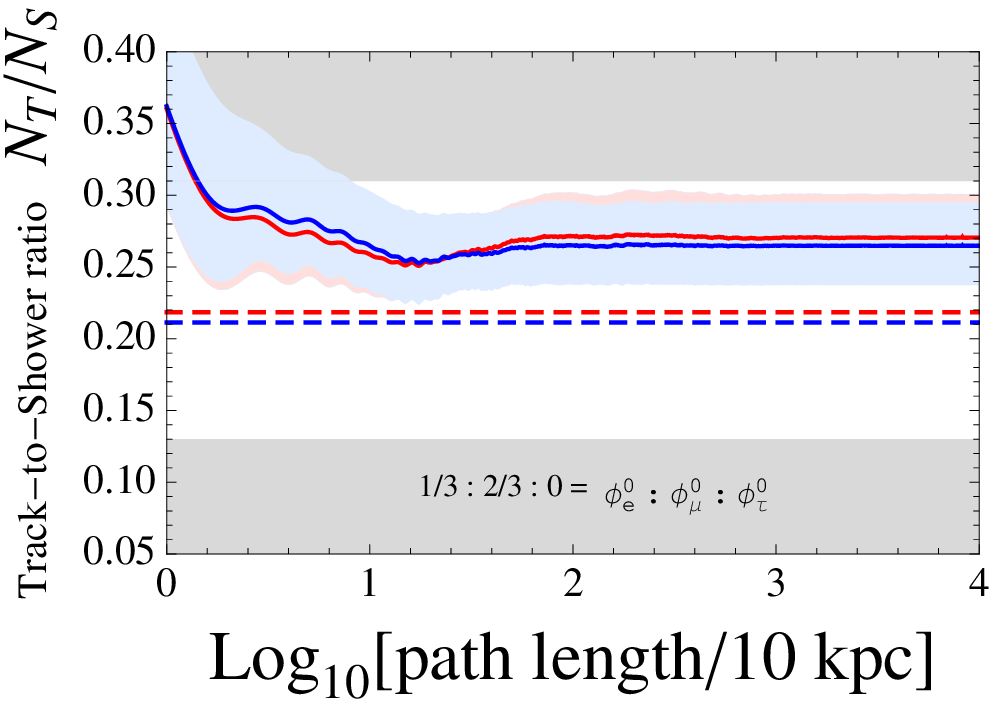,width=8.5cm,angle=0}
\end{minipage}
\hspace*{1.0cm}
\begin{minipage}[h]{7.5cm}
\epsfig{figure=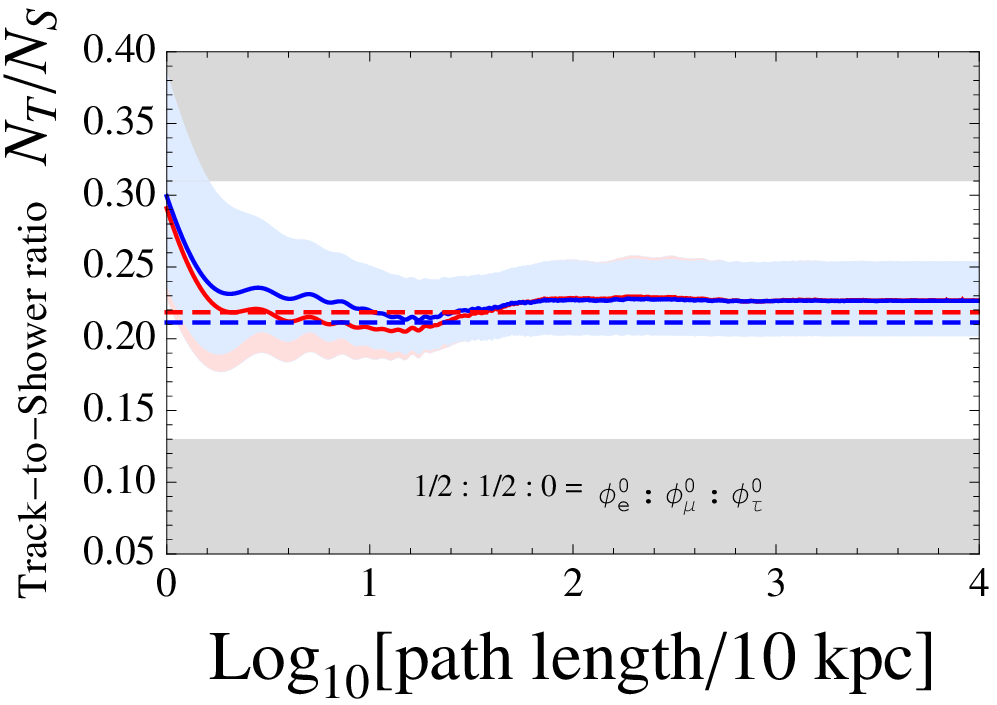,width=8.5cm,angle=0}
\end{minipage}\\
\begin{minipage}[h]{7.5cm}
\epsfig{figure=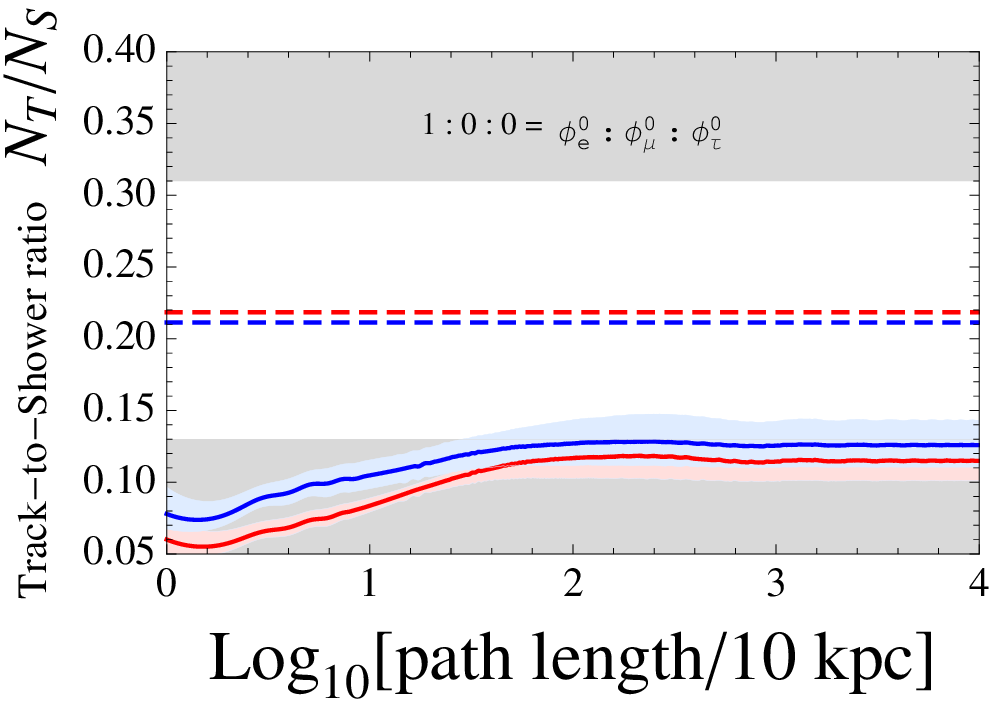,width=8.5cm,angle=0}
\end{minipage}
\hspace*{1.0cm}
\begin{minipage}[h]{7.5cm}
\epsfig{figure=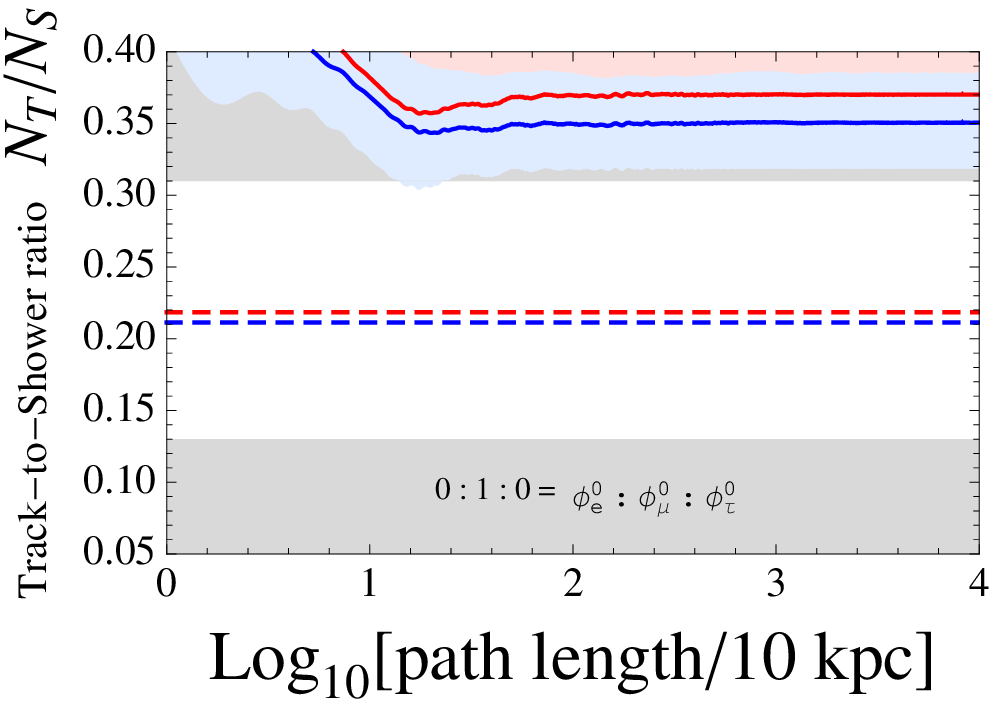,width=8.5cm,angle=0}
\end{minipage}
\caption{\label{FigA6} Plots of the track-to-shower ratio $N_{T}/N_S$ as a function of $L\,(\log_{10}[{\rm path~length/10\,kpc}])$ for NO (red curve) and IO (blue curve) with $\Delta m^{2}_{1}=10^{-15}\,{\rm eV}^2$, $\Delta m^{2}_{2}=10^{-16}\,{\rm eV}^2$ and $\theta_3=1.28$.
Each panel corresponds to the specific initial flavor composition ($\phi^{0}_e: \phi^0_\mu: \phi^0_\tau$) at the source.
  For three neutrino mixing angles and Dirac-type CP phase we take the global fit results in Eq.\,(\ref{mixing_angle}).
Red and blue curved lines correspond to normal and inverted neutrino mass orderings, respectively, for $\omega=2.2$, whereas light red and light blue regions represent the corresponding results for $\omega=1.8-2.6$.
Gray shaded regions are forbidden by $N_T/N_S=0.18^{+0.13}_{-0.05}$ \,\cite{Palladino:2015zua}.}
\end{figure}

We confront  the predictions of $N_T/N_S$ with experimental results 
by taking
$\Delta m^2_1=10^{-15}\,{\rm eV}^2$, $\Delta m^2_2=10^{-16}\,{\rm eV}^2$ and $\theta_3=1.28$ as a benchmark point as well as the best-fit values in Eq.\,(\ref{mixing_angle}) for the neutrino mixing angles and CP phase. 
As can be seen from eqs.(\ref{ntns1},\ref{ak}), 
the tiny mass splittings $\Delta m^2_{k(=1,2)}$ can be searched for, looking at high energy cosmic neutrinos by measuring the track-to-shower ratio $N_T/N_S$ as the function of $L\,(\log_{10}[{\rm path~length}/{\rm 10\,kpc}])$.

In the numerical analysis, we use the spectral index given by $\omega=2.2\pm0.4$\,\cite{Aartsen:2013jdh} and the best-fit values for NO (IO) in Eq.\,(\ref{mixing_angle}).
Fig.\,\ref{FigA6} shows the plots of the track-to-shower ratio $N_{T}/N_S$ as a function of $L\,(\log_{10}[{\rm path~length/10\,kpc}])$ for the neutrino energy $60\,{\rm TeV}\lesssim E_\nu\lesssim3\,{\rm PeV}$ studied in Ref.\,\cite{Palladino:2015zua}.
According to four specific assumptions at each panel for the flavor compositions at the source ($\phi^{0}_e: \phi^0_\mu: \phi^0_\tau$), for $\omega=2.2$, $\Delta m^{2}_{1}=10^{-15}\,{\rm eV}^2$ and $\Delta m^{2}_{2}=10^{-16}\,{\rm eV}^2$ the normal (inverted) mass ordering is presented as the red (blue) curved line, whereas light red and light blue regions represent the corresponding results for $\omega=1.8-2.6$.
Gray shaded regions are forbidden by the measurement $N_T/N_S=0.18^{+0.13}_{-0.05}$ \,\cite{Palladino:2015zua}.
In Fig.\,\ref{FigA6}, we see that the oscillation effect occurs at distance $\lesssim1.5$ Mpc and  it is averagged out at distance beyond 1.5 Mpc.
The predictions of $N_T/N_S$ for the given inputs and the specific initial flavor compositions $1/3: 2/3: 0$ and $1/2:1/2:0$ are consistent with the measurement, whereas
those for the other two initial flavor compositions are disfavored.
In the plots, we draw the horizontal dashed  lines corresponding to the cases without oscillation effects (i.e. the cases for $\Delta m^2_{1,2}=0$.).
The gap between the predictions with and without oscillations is due to the oscillatory term in eq.(\ref{pro_ast}).
Therefore, substantial reduction of uncertainty  in the masurement of $N_T/N_S$ would test not only the model itself but also the oscillation effects induced by pseudo-Dirac nature
of neutrinos.

\section{conclusion}

We have proposed a convincing model containing sterile neutrinos to interpret both SBL neutrino anomalies and high energy neutrino data observed at IceCube in terms of neutrino oscillations.
Different from the so-called $3+1$ model where the PMNS matrix is simply extended to $4\times 4$ unitary matrix as in Refs.\,\cite{Mohapatra:2005wk, white, Gariazzo:2015rra}, the new $4\times 4$ neutrino mixing matrix in our model is parameterized in a way to keep the $3\times3$ PMNS mixing matrix for three active neutrinos unitary.
A characteristic feature of this model is that there are no flavor changing neutral current interactions leading to the conversion of active neutrinos to sterile ones or vice versa.
We have presented new forms of neutrino oscillation probabilities modified by introducing new sterile neutrinos.
In this scenario, there are new mass squared differences ($\Delta m^2_{1,2,3}$, $\Delta S^2_{31}$)
and new mixing angle $\theta_3$ in addition to the standard oscillation parameters associated with
only three active flavor neutrinos.
While $\Delta m^2_{1,2}$ are responsible for astronomical baseline high energy neutrino oscillations, $\Delta m^2_3$ (or $\Delta S^2_{31}$) and $\theta_3$ are usable to interpret SBL neutrino oscillations.

Our model can explain SBL neutrino anomalies in terms of neutrino oscillations at the same level of $3+1$ model.
However,  there still exist small tensions (1) in the reactor and Gallium data, flux normalization and 5 MeV bump observed from $\nu_e\rightarrow \nu_e$ disappearance, (2) between the MiniBooNE data in the region $L/E\gtrsim1.5$ [m/MeV] and the model predictions at $\Delta S^2_{31}=0.6\,{\rm eV}^2$ for $\nu_\mu\rightarrow \nu_e$ appearance for  $\sum m_\nu$ favored by Planck Collaboration (TT+lowP) at $95\%$ CL\,\cite{Moscibrodzka:2016ofe}, and (3) between the Super-Kamiokandes atmospheric neutrino data and the IceCube DeepCore results including the MINOS and MINOS+ data released in Neutrino 2018 .

We have shown that resolution of the reactor antineutrino flux anomaly is required to reduce statistical uncertainties and/or to understand its underlying physics at baselines $L\lesssim500$ m.
In the present model the values of the parameters $\theta_3$ and $\Delta S^2_{31}$ favored by the Planck data of $\sum m_\nu\lesssim0.705$ eV are not conflict with the NEOS and DANSS results\,\cite{Ko:2016owz} which are in a tension with the Gallium and reactor anomalies as in 3+1 model\,\cite{white}.

We have shown that the LSND $\bar{\nu}_e$ appearance data favors probability driven by $\theta_{3}\sim1.28$ with $\Delta S^2_{31}=0.6\,{\rm eV}^2$ satisfying a cosmological bound $\sum m_\nu\lesssim0.705$ eV.
The MiniBooNE excess results are well consistent with the LSND data at $L/E\lesssim1.5$ [m/MeV] while the excess of  two data at $L/E\gtrsim1.5$ [m/MeV] seems to be disfavored by a cosmological bound $\sum m_\nu\lesssim0.705$ eV.
Since the LSND and MiniBooNE data can be interpreted as $\bar{\nu}_\mu \rightarrow \bar{\nu}_e$ oscillation, experimental search for $\bar{\nu}_\mu$ (or $\nu_\mu$) disappearance would test our model in which it could be observed at $L/E\sim1.2$ [m/MeV] with $E_{\nu_\mu}\sim0.5$ GeV and $L=600$ m for $\Delta S^2_{31}=0.6\,{\rm eV}^2$ with $\theta_3=1.28$  in the SBN Program experiment\,\cite{Bass:2017qhe}. 
In addition, we have studied the earth matter effect, and found that  it is negligible for muon neutrinos having energies $\gg1$ GeV when they pass through the interior of the Earth.

Finally, we have found that the existence of light sterile neutrino we consider does not
affect solar neutrino oscillation and thus no constraint on new parameters came out from it.
It has been shown that the Super-Kamiokande's atmospheric neutrino data are consistent with the new $\nu_\mu\leftrightarrow\nu_\tau$ oscillation affected by sterile neutrino, showing  the first oscillation dip
appeared at $\sim 500$km/GeV which is a characteristic feature of this model.
The most recent data of DeepCore\,\cite{Aartsen:2014yll} and MINOS and MINOS+\,\cite{Nu2018} experiments do not show any signature of light sterile neutrinos.
 In addition, we have shown that the probabilities of $P_{\bar{\nu}_\mu\rightarrow\bar{\nu}_\mu}$ and $1-P_{\nu_\mu\rightarrow\nu_\tau}$ versus $L/E$ [km/GeV] have a clear discrepancy with different oscillation signatures of light sterile neutrinos, unlike the expected from the three neutrino standard form.  Such discrepancy could be probed through both the $\nu_\tau$ appearance and the $\nu_\mu$ disappearance experiments in the future.
We have discussed the implications of the very high energy neutrino events detected at IceCube on the probe of the oscillation effcts induced by two pseudo-Dirac mass splittings.

\acknowledgments{We would like to give thanks to E. J. Chun, Xiaojun Bi, S. H. Seo, and Yufeng Li for useful conversations. Y.H. Ahn is supported by the NSFC under Grant No. U1738209. 
}


\end{document}